\documentclass[useAMS,usenatbib]{mn2e}

\usepackage{aas_macros}
\usepackage{amsmath,amssymb}
\usepackage{graphicx}
\usepackage{epstopdf}
\usepackage[normalem]{ulem}
\usepackage[usenames,dvipsnames]{xcolor}
\usepackage{afterpage}
\usepackage{ulem}
\usepackage{pifont}
\usepackage{tikz}

\newcommand{\siiv}{Si~{\small IV} }
\newcommand{\oone}{O~{\small I} }
\newcommand{\cfour}{C~{\small IV} }
\newcommand{\ctwo}{C~{\small II} }
\newcommand{\sfour}{Si~{\small IV} }
\newcommand{\nfive}{N~{\small V} }
\newcommand{\magtwo}{Mg~{\small II} }
\newcommand{\ulas}{ULAS J1120+0641 }
\newcommand{\ula}{ULAS J1120+0641}
\newcommand{\lal}{Lyman-$\alpha$ }
\newcommand{\Lal}{Lyman-$\alpha$ }

\author[Bosman et al.]
  {Sarah E. I. Bosman$^{1,2}$\thanks{seib2@ast.cam.ac.uk}, 
George D. Becker$^{3}$,
Martin G. Haehnelt$^{1}$,
\newauthor
Paul C. Hewett$^{1}$, 
Richard G. McMahon$^{1,2}$,
Daniel J. Mortlock$^{4,5,6}$, 
Chris Simpson$^{7}$,
\newauthor
and Bram P. Venemans$^{8}$ \\
  $^1$Institute of Astronomy, University of Cambridge, Madingley Road,
Cambridge CB3 0HA, U.K.\\
  $^2$Kavli Institute for Cosmology, University of Cambridge, Madingley Road,
Cambridge CB3 0HA, U.K. \\
  $^3$Department of Physics \& Astronomy, University of California, Riverside, 900 University Avenue, Riverside, CA, 92521, USA \\
  $^4$Astrophysics Group, Imperial College London, Blackett Laboratory, Prince Consort Road, London SW7 2AZ, U.K. \\
  $^5$Department of Mathematics, Imperial College London, London SW7 2AZ, U.K. \\
  $^6$Department of Astronomy, Stockholm University, Albanova, SE-10691 Stockholm, Sweden \\
  $^7$Gemini Observatory, Northern Operations Center, 670 N.~A`oh\={o}k\={u} Place, Hilo, HI 96720-2700, USA \\
  $^8$Max-Planck Institute for Astronomy, K\"onigstuhl 17, 69117
Heidelberg, Germany \\} 

\title[ULAS J1120+0641: Metals during Reionisation]{A deep search for metals near redshift 7: the line-of-sight towards ULAS J1120+0641}
\date{}
\begin{document}
\maketitle

\begin{abstract}
We present a search for metal absorption line systems at the highest redshifts to date using a deep (30h) VLT/X-Shooter spectrum of the $z=7.084$ quasi-stellar object (QSO) \ula. 
We detect seven intervening systems at $z > 5.5$, with the highest-redshift system being a \cfour absorber at $z = 6.51$.
We find tentative evidence that the mass density of \cfour remains flat or declines with redshift at $z < 6$, while the number density of \ctwo systems remains relatively flat over $5 < z < 7$.  These trends are 
broadly consistent with 
 models of chemical enrichment by star formation-driven winds that include
a softening of the ultraviolet background towards higher redshifts.
We find a larger number of 
weak ($W_\text{rest}<0.3$~\AA) \magtwo systems over $5.9 < z < 7.0$ than predicted by a power-law fit to the number density of stronger systems. This is consistent with trends in the number density of weak \magtwo systems at $z \lesssim 2.5$, and suggests that the mechanisms that create these absorbers are already in place at $z \sim 7$. 
Finally, we investigate the associated narrow Si~{\small IV}, C~{\small IV}, and N~{\small V} absorbers located near the QSO redshift, and find that at least one component shows evidence of partial covering of the continuum source.
\end{abstract}


\section{Introduction}\label{sec:introduction} 

\begin{figure*}
\includegraphics[width=\textwidth]{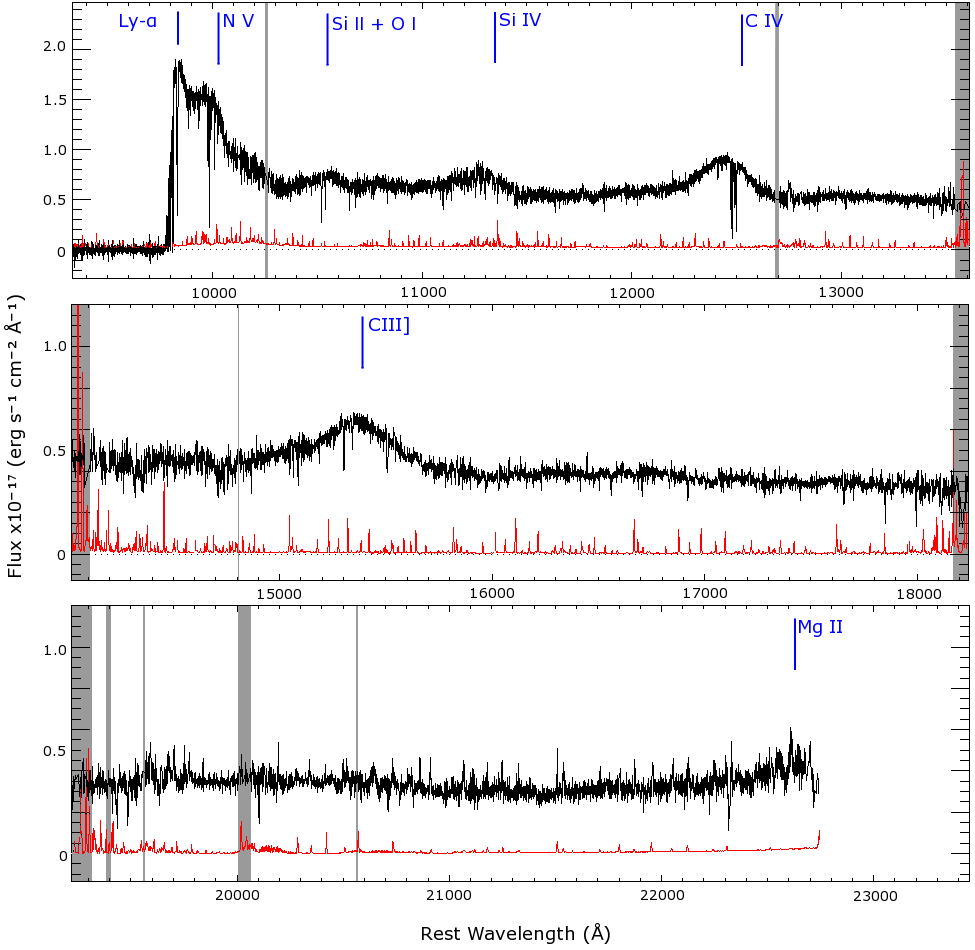}
\caption{Our 30-hour VLT/X-SHOOTER spectrum of ULAS J1120+0641. Prominent emission lines are marked assuming a systemic redshift of $z=7.084$. The spectrum is plotted using 10  km $\text{s}^{-1}$ pixels.  Areas of overlap between the arms of the spectrograph, as well as regions affected by sky residuals with SNR too poor to allow for the metal search to be conducted are highlighted in grey. To demonstrate the mean SNR, individual pixels affected by skylines have not been plotted. The error spectrum is binned (but not rescaled) in bins of 50 km $\text{s}^{-1}$ to avoid crowding the plot with skylines. Some absorption systems can be seen, for instance near the \cfour broad emission line. } 
\label{fig:spectrum}
\end{figure*}

High-redshift quasi-stellar objects (QSOs) are powerful and versatile probes for studying the both intergalactic medium (IGM) and the circum-galactic media around galaxies near the epoch of reionisation. 
The study of \lal transmission along the line-of-sight to QSOs has revealed a rapidly evolving IGM at $z\sim6$ (e.g., \citealt{Fan06}, \citealt{Becker15}), suggesting that hydrogen reionisation may be in its final stages near that redshift (e.g., \citealt{Gnedin06}, \citealt{Mesinger10}, \citealt{Gnedin16}). 
Meanwhile, QSO near zones offer a valuable probe of the ultraviolet background (UVB) up to at least $z\sim 6$ (\citealt{Wyithe14}, \citealt{Bolton07-prox}, \citealt{Maselli09}, \citealt{Carilli10}). 
The proximity zone of the highest-redshift known QSO, \ula\ ($z=7.084$), has even provided hints of a partially neutral IGM and/or chemically pristine circum-galactic gas at $z \sim 7$ (\citealt{Mortlock11}, \citealt{Simcoe12}, \citealt{Bolton11}, \citealt{Greig16}; but see \citealt{Bosman15}, \citealt{Keating15}).  

Metal absorption lines tend to trace the metal-enriched gas located around galaxies, and thus probe a different mass and density regime than the \Lal forest and near zones.
Metal lines provide insight into star formation and feedback processes, and also
offer a means to study galaxies that are too faint to detect in emission. 
In addition, the elemental abundances in metal absorbers provide crucial information on the nature of the earliest stellar populations.

A number of studies have have now traced metal enrichment out to $z \sim 6$ using both highly ionized (C {\small{IV}}, Si {\small{IV}}) and low-ionization (e.g., \ctwo, O {\small{I}}, and \magtwo) species (\citealt{RyanWeber06, Ryanweber09};\citealt{Simcoe06, Simcoe11}; \citealt{Becker09, Becker11}; \citealt{Matejek12, Matejek13}, \citealt{Dodorico13}).  For \magtwo, the redshift frontier has recently been pushed back to $z =7$ \citep{Chen16}.  Meanwhile, numerical simulations have explored the effects of varying star formation histories, early galaxy feedback mechanisms, and the effect of a declining ultraviolet background (UVB) on the number density and ionisation state of metal absorbers (e.g., \citealt{Oppenheimer09}, \citealt{Finlator15}, \citealt{Keating16}). A recent review of this subject can be found in \cite{Beckerreview}.  

The incidence rate of both highly ionized and low-ionization species is observed to evolve with redshift, although the rate of the evolution is still somewhat unclear.  For example, the incidence rate of \cfour appears to decrease from $z=1.5$ to $z=3.5$ and may decrease faster with redshift above $z=4.5$, falling by a factor of $\sim10$ over the entire range \citep{Dodorico13}.  This has been interpreted as the result of ongoing carbon enrichment in the vicinity of galaxies due to outflows (e.g., \citealt{Oppenheimer06}) as well as a possible softening of the UVB towards higher redshift, which impacts the ionization state of carbon (\citealt{Oppenheimer09}; \citealt{Keating14}).

The abundance of low-ionisation metals such as \ctwo and \oone is also observed to decrease with redshift over $1.5<z<5.5$; however, there have been indications that the volume density of \oone systems might be stabilising or even increasing at $z\gtrsim6$ \citep{Becker06}. This trend may be linked to an evolution in the UVB and in the physical densities of metal-enriched gas in ways that tend to favour lower ionisation states at higher redshifts, even while the overall metal enrichment is lower.  Simulations have supported this view (e.g. \citealt{Finlator15, Finlator16}; \citealt{Oppenheimer09}), and predict that \ctwo systems might become equally or more abundant than \cfour systems at $z\gtrsim8$.

Mg~{\small II} has been used extensively to probe metal-enriched gas at lower redshifts ($z \lesssim 2.5$) (e.g., \citealt{Weymann79}; \citealt{Churchill99}; \citealt{Nestor05}; \citealt{Lundgren09}; \citealt{Weiner09}; \citealt{Chen10}, \citealt{Menard11}; \citealt{Kacprzak11,Kacprzak11a}; \citealp{Churchill13a,Churchill13b}; \citealt{Mathes17}). At higher redshifts, infrared surveys by \citep{Matejek12} and \citep{Chen16} have revealed that strong (rest-frame equivalent width $W >$ 1.0 \ \AA) absorbers decline with redshift at a rate similar to the global star formation rate, whereas weaker (0.3 $< W <$ 1.0 \ \AA) systems show little or no decline over $0.4< z < 7$. These observations, along with comparative studies of absorber and galaxy properties, have led to the hypothesis that strong \magtwo systems trace transient phenomena, such as metal-enriched outflows, that track the global star-formation rate,
 while weak  \magtwo systems may be more often associated with inflows, or arise from the fragments of older star-formation-driven winds (see summaries in \citealt{Kacprzak11a}; \citealt{Matejek13}; and \citealt{Mathes17}).

The goal of this paper is to provide a sensitive survey for metal lines out to $z = 7$.
Towards this aim, we have acquired a deep Very Large Telescope (VLT) X-Shooter spectrum of \ulas (hereafter J1120; \citealt{Mortlock11}) which, at a redshift of $z=7.0842 \pm 0.0004$, \citep{Venemans12}, is curently the most distant known QSO.
In terms of its UV continuum, it is also one of the most luminous known QSOs at $z>6.0$, making it an excellent target for spectroscopic follow-up, and a powerful probe of metal absorbers beyond redshift six.  For \cfour and \ctwo these are the first observations of intervening metal lines out to $z=7$, while for \magtwo we probe considerably lower equivalent widths than previous studies.

The remainder of the paper is organised as follows.  We describe our X-Shooter spectrum of J1120 in Section~\ref{sec:data}, and our methodology for identifying and measuring intervening metal absorption lines in Section~\ref{sec:method}, which also outlines the analysis techniques used for extracting number densities, column density distribution functions, and cosmic mass fractions. Our results are presented in Section~\ref{sec:results}, in which the implications for C {\small{IV}}, Mg {\small{II}} and C {\small{II}} are analysed in turn and compared to results at lower redshifts and predictions from numerical simulations. The last part of Section~\ref{sec:results} presents evidence for partial covering of the QSO line-of-sight by associated absorbers in C {\small{IV}} and N {\small{V}}.  A summary of our results is given in Section~\ref{sec:summary}.  Throughout this paper we assume a flat cosmology with $[\Omega_M, \Omega_k, \Omega_\Lambda, h] = [0.3, 0, 0.7, 0.7]$ and equivalent widths are quoted in the rest frame unless explicitly stated. When quoting uncertainties, we give Bayesian 68\% credible intervals unless explicitly stated otherwise.

\begin{figure}
\includegraphics[width=\columnwidth]{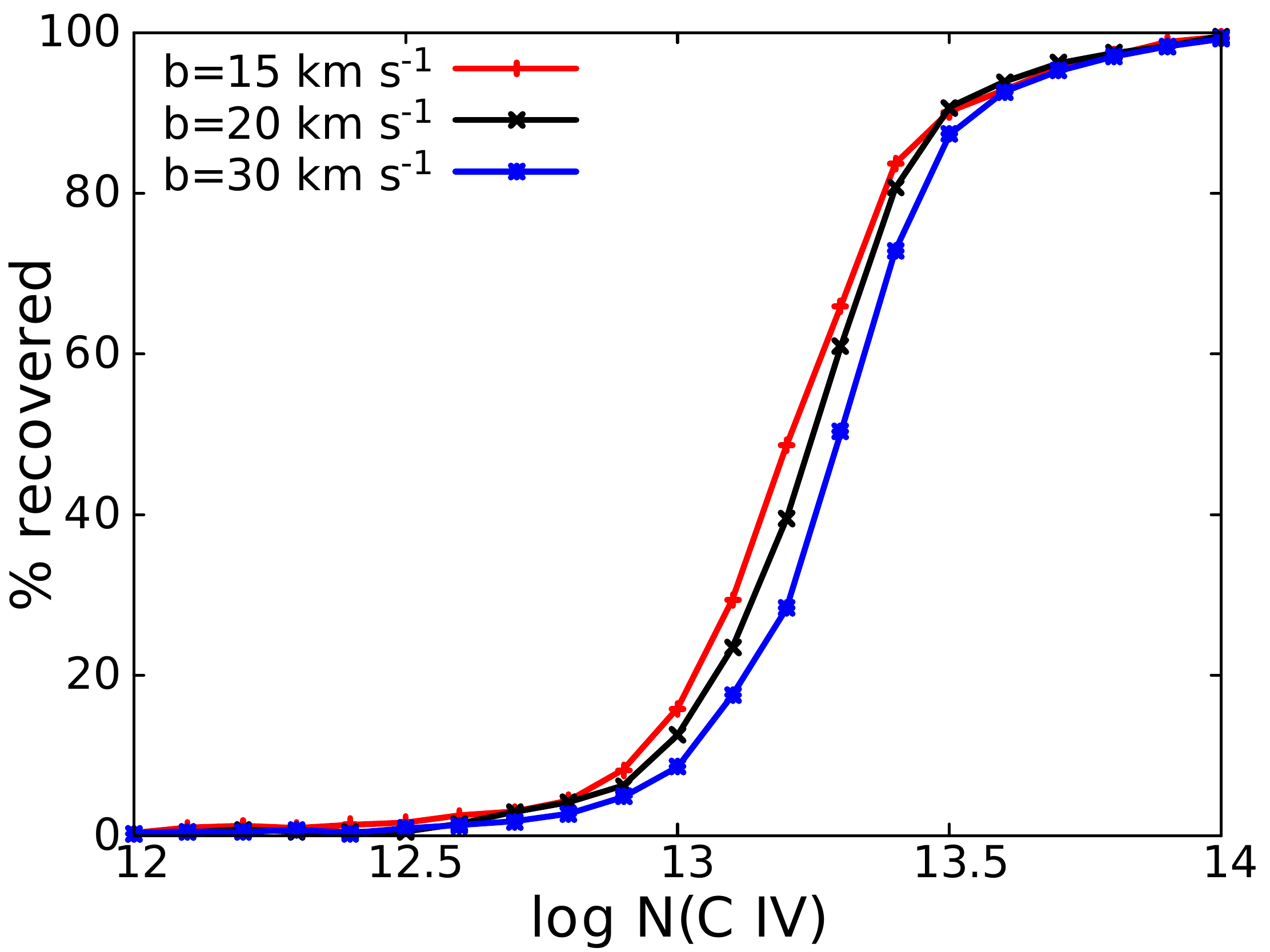}
\caption{\cfour completeness as a function of column density.  Each point was obtained by inserting an artificial \cfour doublet at a randomly chosen redshift over $5.3<z<7.0$ (as plotted in Figure~\ref{fig:summary}) 1000 times, and attempting recovery with the automated line detection algorithm described in Section~\ref{sec:method}.  The `+', `x', and `$*$' symbols denote results for $b = 15$, 20, and 30 km s$^{-1}$, respectively. Completeness estimates over $6.2<z<7.0$, over which the fitting is done, are similar.
\label{fig:cfour_comp}}
\end{figure}

\begin{figure}
\includegraphics[width=\columnwidth]{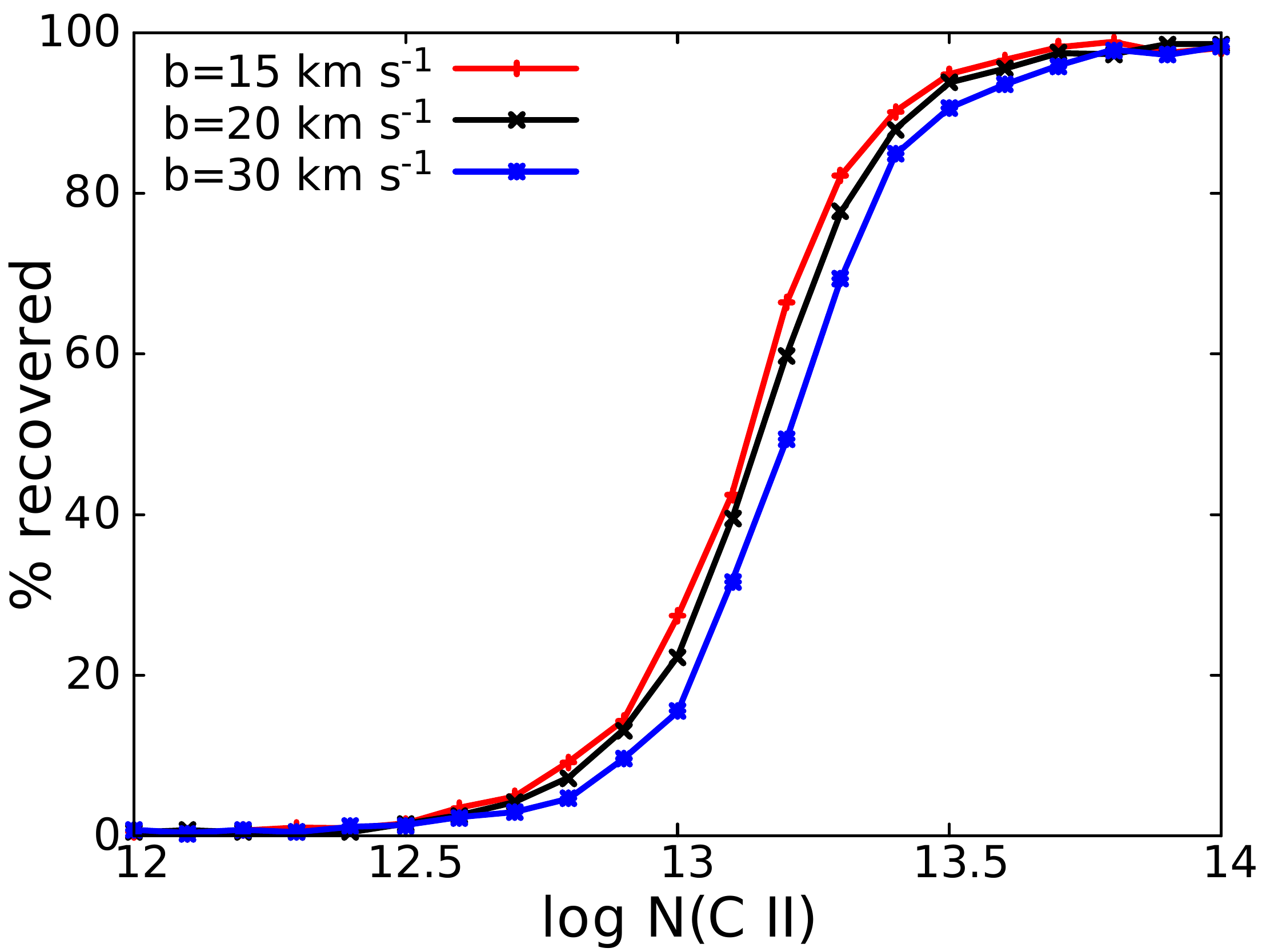}
\caption{\ctwo completeness as a function of column density. Range searched extends over $6.3<z<7.0$. Symbols are as in Figure~\ref{fig:cfour_comp}.} 
\label{fig:cii_comp}
\end{figure}

\begin{figure}
\includegraphics[width=\columnwidth]{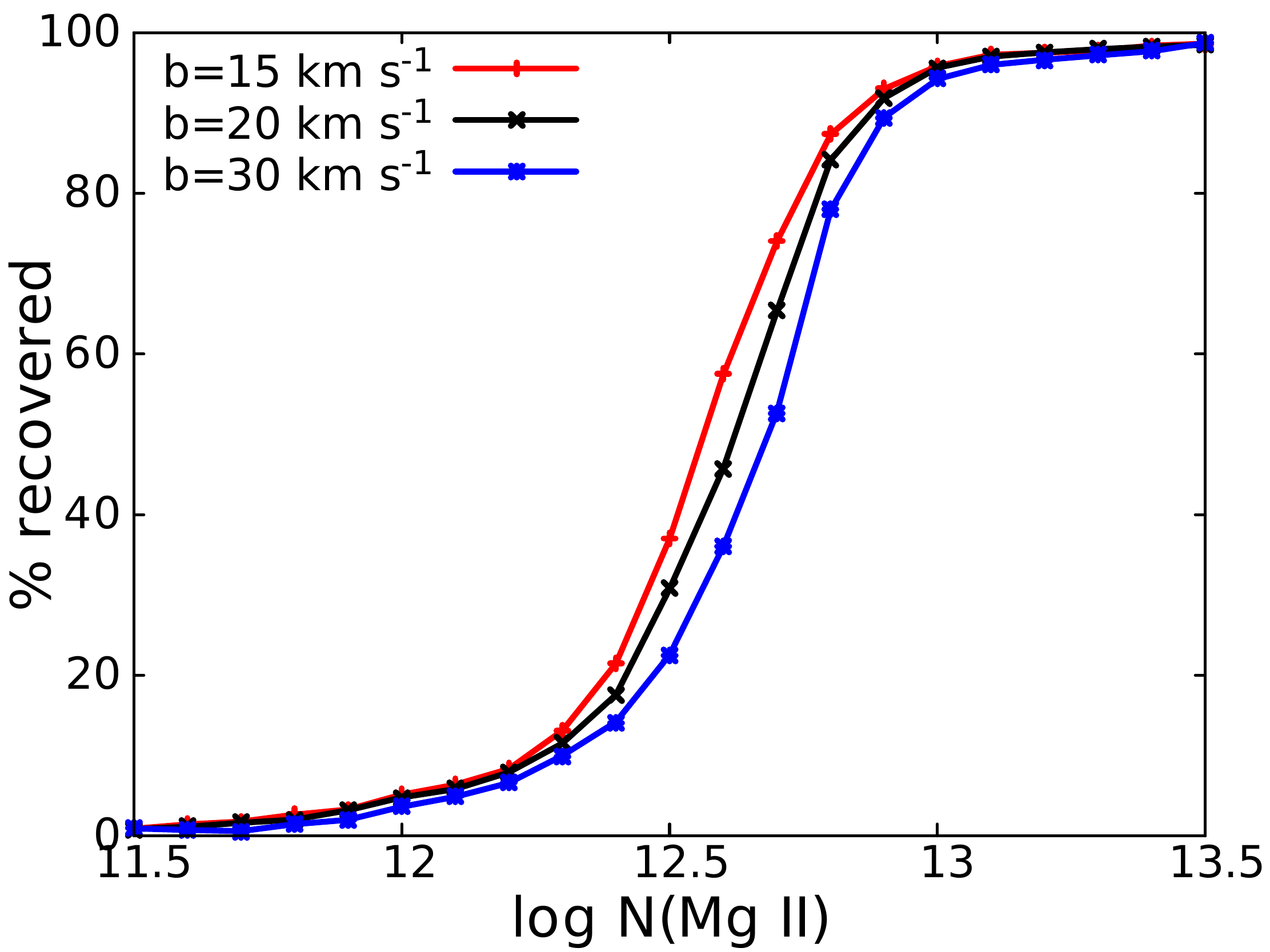}
\caption{\magtwo completeness as a function of column density. Range searched extends over $5.9<z<7.0$. Symbols are as in Figure~\ref{fig:cfour_comp}.} 
\label{fig:mgii_comp}
\end{figure}

\section{Data}\label{sec:data}

We obtained a 30-hour VLT/X-Shooter spectrum of J1120 in observations spanning March 2011 to April 2014.\footnote{ESO programmes 286.A-5025(A), 089.A-0814(A), and 093.A-0707(A).}  The data were reduce using a suite of custom {\sc idl} (Interactive Data Language\footnote{http://www.exelisvis.com}) routines.  Individual exposures were flat-fielded and sky-subtracted using the optimal method described by \citet{Kelson03}.  Relative flux calibration was applied to the two-dimensional frames using response curves derived from standard stars.  A single one-dimensional spectrum using 10 km s$^{-1}$ bins was then optimally extracted from all exposures simultaneously.   Telluric correction was performed using SkyCalc\footnote{http://www.eso.org/sci/software/pipelines/skytools/} atmospheric transmission models.  Absolute flux calibration was performed by scaling the corrected spectrum to match the VLT/FORS2 and GNIRS spectra of J1120 obtained by \cite{Mortlock11}.  For the absorption lines analysis, the region redward of the the Ly$\alpha$ forest was normalized using a slowly-varying spline fit.  Slit widths of 0\farcs9 were used in the VIS and NIR arms, giving nominal resolutions of $R = 7450$ and 5300, respectively.  Inspection of the telluric absorption lines, however, indicated that the true mean resolution was somewhat better, $R \simeq 10000$ and 7000, consistent with a typical seeing ${\rm FWHM} \simeq $ 0\farcs7.

The combined, flux-calibrated spectrum is shown in Figure~\ref{fig:spectrum}.   
The continuum signal-to-noise ratio spans $S/N \sim 10$ to 50 per 10 km\,s$^{-1}$ pixel outside of regions strong affected by the atmosphere.  Notably, the residuals from strong sky emission lines tend to exceed the estimates from the formal error array.  This appears to be at least partly due to the fact that the projected slit width in the raw frames changes very slightly from one end of the slit to the other.  This makes it difficult to model the sky purely as a function of wavelength, and requires an additional fit in the spatial direction.  For this we used a variable low-order polynomial; however, non-Poisson residuals still remained for strong lines.  We therefore treat regions affected by skylines with caution in our analysis.

\section{Search Method}\label{sec:method}

\begin{figure*}
\includegraphics[width=\textwidth]{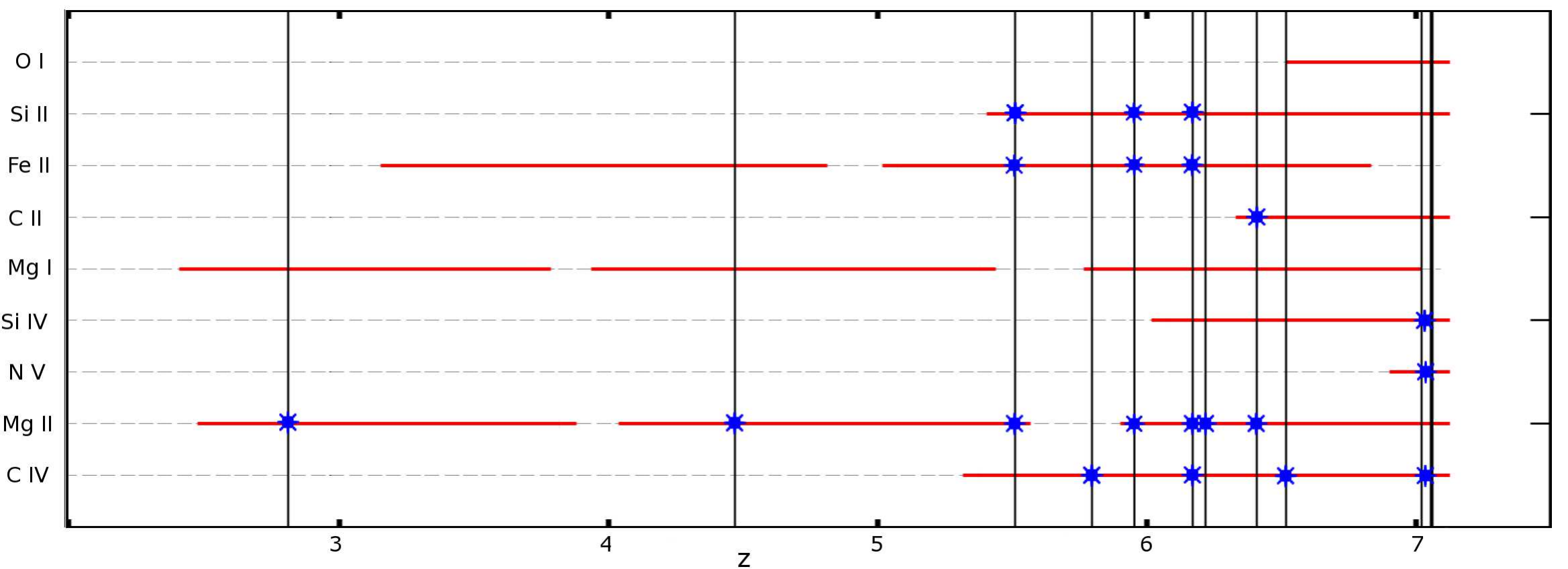}
\caption{A graphic summary of our survey results.  The redshift ranges over which we searched for different ions are shown by horizontal solid lines.  The redshifts of detected systems are indicated by vertical solid lines, with detected species marked by blue stars. The emission redshift of the QSO is indicated by a thick vertical line at $z=7.084$. Nine intervening systems are found, along with three associated systems within 3000 km $\text{s}^{-1}$ of the object (shown on the Figure as a single line).}
\label{fig:summary}
\end{figure*}

Unlike searches for metal absorption lines at low redshift, where the \lal forest provides a guide as to the redshifts of intervening systems, the onset of nearly complete Ly$\alpha$ absorption at $z\gtrsim5.5$ means we are forced to rely solely on the metal lines themselves for identification. For this reason we use a modified version of the detection technique used by \citet{Becker09, Becker11}, wherein multiple lines are identified with an absorber at a single redshift based on their relative wavelength and optical depth ratios.  This process is straightforward for doublets such as \cfour $\lambda\lambda 1548,1551$ and \magtwo $\lambda\lambda 2796,2804$, multiple lines of the same species (as for Fe~{\small II}), and single lines that commonly arise from the same absorber (e.g., \ctwo $\lambda 1334$ and O~{\small I} $\lambda 1302$).  The final search consisted of the ions listed in Table~\ref{tab:ions}.  The search window extended from $\lambda = $ 22750 \ \AA \ down to 9838 \ \AA\  (i.e., the onset of the Ly$\alpha$ forest). Wavelengths affected by strong telluric absorption were masked.  The redshift range over which we searched for each ion is shown in Figure~\ref{fig:summary}.

\begin{table}
\caption{List of the major ion lines included in the search.}
\centering
\begin{tabular}{c c}
\hline
Ion & $\lambda$/\AA\\
\hline
N~{\small V} &  1238.82, 1242.80  \\
O~{\small I } & 1302.16  \\
C~{\small II} & 1334.53  \\
Si~{\small IV} & 1393.76, 1402.77 \\
\cfour & 1548.20, 1550.77  \\
Si~{\small II} & 1526.71  \\
Al~{\small III} & 1854.71, 1862.79  \\
Fe~{\small II} & 2344.21, 2382.76, 2586.65, 2600.17 \\
Mg~{\small II} & 2796.35, 2803.53  \\
Mg~{\small I} & 2852.96\\
\hline
\end{tabular}
\label{tab:ions}
\end{table}

\begin{table}
\caption{List of detection thresholds for the species of interest. Throughout the paper we make use of the 15 km $\text{s}^{-1}$ values for all saturated systems.}
\centering
\begin{tabular}{c c c}
\hline
Ion & $b$/km $\text{s}^{-1}$ & $\log (N_\text{min} / \text{cm}^{-2})$ \\
\hline
\cfour & 15 & 13.1 \\
 & 30 & 13.2  \\
Mg~{\small II} & 15 & 12.5 \\
 & 30 & 12.7 \\
C~{\small II} & 15 & 13.0  \\
 & 30 & 13.2 \\
Si~{\small IV} & 15 & 12.6\\
 & 30 & 12.7 \\
Mg~{\small I} & 30 & 12.3 \\
Si~{\small II} & 30 & 13.2 \\
\hline
\end{tabular}
\label{tab:thresholds}
\end{table}

We used a two-step process to perform the initial line detection.  First, all lines with $\tau\gtrsim0.3$ were identified visually.  A second, automatic line identification procedure was then applied. 
The automatic algorithm used an inverted Gaussian template, with two free parameters for the depth and width. 
At regularly spaced intervals in velocity $\Delta v = 20$ km $\text{s}^{-1}$, this template was fit to a small region of the spectrum, iteratively rejecting pixels that exceeding a 2$\sigma$ clipping threshold.

To aid identification, preliminary column densities were derived using the apparent optical depth formula \citep{Savage91} in which the optical depth $\tau$ is related to the observed flux intensity, $F$, and the continuum intensity, $F_0$, as
\begin{equation}  
\tau = - \ln\left( F / F_0 \right) \, .
\label{eq:opt}
\end{equation}
The ionic column density, $N_i$, for a pixel of optical depth $\tau_i$ is then derived as
\begin{equation}
N_i = \frac{m_\text{e} c}{\pi e^2} \frac{\tau_i}{ f \lambda_0 } \, ,
\label{eq:aod}
\end{equation}
where $f$ is the oscillator strength of the relevant ion transition, $\lambda_0$ is its wavelength, $e$ the electron charge and $m_e$ is the electron mass. 
Column densities are quoted in units of $\text{cm}^{-2}$ throughout. 

The column density thresholds for detection, given in Table~\ref{tab:thresholds}, were chosen so that the number of false detections, following the  criteria described below, was either zero or had reached a baseline level driven by bad pixels, where a candidate detection was easily eliminated by visual inspection.  These thresholds typically corresponds to column densities where we are $\sim$30 per cent complete.  
In addition, we required the flux decrement across the selected absorption features to be significant to at least 5 $\sigma$ based on the noise array.

For positively identified systems, final column density and Doppler $b$ parameters (in km $\text{s}^{-1}$) were obtained by fitting Voigt profiles using the fitting program {\tt{vpfit}} \citep{vpfit}. This also allowed us to introduce a variable power-law correction to the continuum for each line.  
The detected systems were all
initially fit using a single velocity component. Three systems
show evidence of more complex kinematics, and for these we
also performed multiple component fits, as shown in the Appendix. The multiple component fits were typically poorly
constrained due to the noise levels and resolution of the data;
however, in all cases the sum value of the column densities of
the individual components falls within the error bounds for
the single component fits, which are given in Table~\ref{tab:summary}.
 Where required, we computed upper limits for the column density of the undetected ions in identified intervening systems by inserting increasingly strong artificial lines near the corresponding redshift. An offset equal to the Doppler parameter of the detected ions in the system was chosen to mitigate the effect of potential absorption lines just below the detection threshold.
The upper limit corresponds to the weakest injected line which would still be detected independently from associated ions, using the same detection criteria
near  that wavelength.

\begin{table*}
\caption{Intervening systems along the ULAS J1120+0641 line-of-sight. A dash `--' in the column density column indicates that the ion would occur outside of the range probed by our spectrum. Upper limits are given for non-detections. Doppler parameters of less than 15 km $\text{s}^{-1}$ indicate that the absorption feature is unresolved with X-Shooter.}
\centering
\makebox[2.0\textwidth][l]{ \hspace{-4em}
\begin{tabular}{c c c c c c c c c}
\hline
$z_\text{abs}$ & log $N_{\text{C IV}}/\text{cm}^{-2}$ & $b_\text{C IV}$ &log  $N_{\text{Mg II}}/\text{cm}^{-2}$ &$b_\text{Mg II}$ &log  $N_{\text{Mg I}}/\text{cm}^{-2}$ &log  $N_{\text{C II}}/\text{cm}^{-2}$ & log $N_{\text{Fe II}}/\text{cm}^{-2}$ & log $N_{\text{Si II}}/\text{cm}^{-2}$ \\
\hline
6.51511 & 13.25$^a$ $\pm$ 0.06 & 12 $\pm$ 8 & $<13.0$ &  & $<12.0$ & $<12.3$ & $<12.7$ & $<11.9$ \\
6.40671 & $<13.1$ &  &12.8 $\pm$ 0.2 &9 $\pm$ 6& $<11.6$ & 13.4 $\pm$ 0.4 & $<12.1$ & $<12.0$ \\
6.21845 & $<13.2$ &  &12.57 $\pm$ 0.07 & 12 $\pm$ 9& $<11.7$ & -- & $<12.4$ & $<11.9$ \\
6.1711 & 13.67 $\pm$ 0.03& 57 $\pm$ 6 & 13.4 $\pm$ 0.6 & 19 $\pm$ 5& $<11.8$ & -- & 12.7 $\pm$ 0.2 & 13.2 $\pm$ 0.4 \\
5.9507 & $<13.2$ &  & 13.1 $\pm$ 0.1 & 42 $\pm$ 3 & $<12.0$ &  -- & 12.8 $\pm$ 0.1 & 13.2 $\pm$ 0.2\\
5.79539 & 13.97 $\pm$ 0.03& 40 $\pm$ 2& -- & -- & $<12.0$ & -- & $<11.5$ & $<13.1$ \\
5.50793 & $<12.7$ &  & 13.37 $\pm$ 0.04 & 49 $\pm$ 5& -- & -- & 13.1 $\pm$ 0.1 & 13.5 $\pm$ 0.1\\
4.47260 & -- & -- & 12.89 $\pm$ 0.04 & 13 $\pm$ 2& $<11.6$ & -- & $<11.7$ & -- \\
2.80961 & -- & -- & 12.81 $\pm$ 0.06 & 12 $\pm$ 4& $<11.8$ & -- & -- & -- \\
\hline
\end{tabular}}
\begin{flushleft}
$^a$ This column density is for the Voigt profile fit to the region in Figure~\ref{fig:cfour_z6p51} that is uncontaminated by skyline residuals in both \cfour transitions.  Integrating the apparent optical depths over the full $\lambda$1548 profile gives $\log{N_\text{C IV} / {\rm cm^{-2}}} = 13.46 \pm 0.04$.
\end{flushleft}
\label{tab:summary}
\end{table*}

Detection completeness was evaluated by inserting mock absorbers of each metallic species into the spectrum and attempting to recover them over a range of velocity width parameters and column densities.  
The redshift ranges probed are the ones shown in Figure~\ref{fig:summary} and vary between ions; the Doppler parameters tested are $b= 15, 20, 30$ km $\text{s}^{-1}$ for the most common species \cfour, \magtwo, \ctwo and \siiv, and $b=30$ km $\text{s}^{-1}$ for less common species. For each combination of $b$ and $N$, a redshift from the probed range is chosen at random, then an absorption line with those parameters is injected in the spectrum. The search algorithm is then run to attempt to recover the artificial line at $5\sigma$ significance based on the error array and at a threshold higher than that of the false positives (see next paragraph). 
Example results are shown in Figures 2, 3, and 4 for \cfour, \ctwo, and \magtwo, respectively.  We find that we are able to detect $>30$ per cent of \cfour absorbers of column density $\log (N / \text{cm}^{-2}) = 13.1$ and $>95$ per cent at $\log (N / \text{cm}^{-2}) = 13.7$, for $b=15$ km $\text{s}^{-1}$.
Due to the higher oscillator strength of the ion, we are sensitive to 30\% of \sfour systems with $\log (N / \text{cm}^{-2}) = 12.7$ \ and $>95$ per cent of those with $\log (N / \text{cm}^{-2}) \geq 13.3$ \ with Doppler parameter $b = 15$ km $\text{s}^{-1}$. 
Similarly, we are able to detect $>30\%$ per cent of \magtwo absorbers of column density $\log (N / \text{cm}^{-2}) = 12.5$ and $>95$ per cent at $\log (N / \text{cm}^{-2}) = 12.9$ for $b=15$ km $\text{s}^{-1}$.

To mitigate contamination from false positives, we chose a column density threshold for each species above which detections are considered to be reliable. We determined this threshold by estimating the number of false positives, for a range of ions and column densities, by 
(i) inserting artificial doublets with incorrect optical depths ratios, based on relative oscillator strengths of the transitions and re-running the detection algorithm to check that no such systems are picked up as valid detections;
(ii) inserting artificial doublets with slightly ($\sim 20\%$) incorrect velocity spacing to check the code's sensitivity to spurious interlopers, (iii) searching for doublets with incorrect velocity spacing which should not exist, to check that the code does not pick up chance alignments of noise fluctuations, and (iv) inverting the sign of the Gaussian template to look for spurious lines in emission, and checking that no such features are detected above the chosen column density thresholds.
Finally, we ran the code on wavelength ranges visibly devoid of absorbers to check the results were in good agreement with visual inspection.

\section{Results}\label{sec:results}

\subsection{Overview}\label{sec:overview}

In total we detect twelve absorption systems.  Nine of these are intervening, and seven of these intervening systems are located at $z > 5.5$.  We identify three associated absorbers within 3000 km $\text{s}^{-1}$ of the QSO redshift.  We do not include these three systems in out main sample; however, two of the absorbers, which appear to be associated with the QSO itself, are analysed further in Section~\ref{sec:associated}.  A summary of the absorber properties is given in Table~\ref{tab:summary}.

Plots of the intervening systems can be found in the appendix. The two systems at $z<5.5$ are detected through \magtwo only, with tight upper limits on Mg~{\small I}. No other ions are covered over this redshift range. Meanwhile the seven $z > 5.5$ objects display a wide range of ionic ratios, with five of them displaying \magtwo and three of them displaying \cfour up to a redshift of $z=6.515$, currently the highest redshift detection of an intervening metal absorber along a QSO line-of-sight. Fe~{\small II} and Si~{\small II} are found in systems located at $z=5.508$, $5.950$, and $z=6.1714$, while C~{\small II} is detected with \magtwo at $z=6.406$. Notably, our spectral coverage would allow us to detect \sfour systems located at $z > 6.0$, but none are detected.

Finally, no low-ionisation absorbers are seen in the redshift interval where O~{\small I} would be detected. 
We therefore find no apparent overabundance of O~{\small I} along this line-of-sight, despite indications that such systems might become more common at higher redshifts \citep{Becker11}. This may be due to scatter between lines-of-sight and a narrow visibility interval ($6.6 < z <7.0$, corresponding to an absorption path length $\Delta X = 2.0$). On the other hand, we do detect a significant number of \magtwo systems at $z > 5.9$, as discussed below.

\subsection{Statistics}\label{sec:stats}

We compute a range of standard statistics for different metal species.  The number density of absorbers is computed alternately per unit redshift, $\Delta z$, and per unit absorption path length, $\Delta X$, where
\begin{equation}
 X(z) = \int_{0}^{z} (1+z')^2 \frac{H_0}{H(z')} \text{d}z' 
\label{eq:abs}
\end{equation}
\citep{Bahcall69}.  The column density distribution function (CDDF), 
\begin{equation}
f(N) = \frac{\partial^2 n}{\partial N \partial X}
\label{eq:cddf}
\end{equation}
can be integrated to obtain the cosmic mass density for a species, usually expressed as a fraction of the critical mass density, $\rho_{\rm crit} = 1.88 \times 10^{-29} h^2 \text{\ g cm}^{-3}$, as 
\begin{equation}
\Omega_{\rm ion} = \frac{H_0 m_\text{ion}}{c \rho_{\text{crit}}} \int \text{d}N N f(N) \, .
\label{eq:omega}
\end{equation}
In practice the mass fraction is computed over a limited range of column densities.  We correct for completeness when computing these quantities, as described below.

\subsection{\cfour}\label{sec:civ}

Previous \cfour studies have shown a decline in the comoving mass density of \cfour between $z \sim 1.5$ and $z \sim 4$ (e.g., \citealt{Dodorico10}, \citealt{Boksenberg15}), with a possible acceleration of the decline from $z \sim 4.5$ to $z \sim 5.5$ (\citealt{Becker09}, \citealt{Ryanweber09}, \citealt{Simcoe11}, \citealt{Dodorico13}). The column density distribution function of absorbers is normally described by a power law whose slope is roughly consistent across $1.5<z<5.5$; however, the normalisation of the power law falls by a factor of $\sim$10 with redshift over the same range \citep{Dodorico13}.  On the modeling side, this has been interpreted as the result of ongoing carbon enrichment in the vicinity of the host galaxies, coupled with a softening of the UVB towards higher redshifts (e.g., \citealt{Oppenheimer09}; \citealt{Oppenheimer09}; \citealt{Finlator15, Finlator16}).

We use the J1120 line-of-sight to assess whether the observed decline in $\Omega_{\text{C IV}}$ continues at $z>6.2$, the highest redshift probed by earlier surveys. Our \cfour search above this redshift extends over $6.2<z<7.0$, corresponding to $\Delta X = 4.0$.  We find only one \cfour absorber in this range, at $z=6.515$.  The blue edge of the $\lambda$1551 profile is affected by skyline residuals (Figure~\ref{fig:cfour_z6p51}).  A Voigt profile fit to the the velocity range uncontaminated in both doublet transitions gives $\log N=13.25 \pm0.06$.  Integrating the optical depths over the full apparent $\lambda$1548 profile and applying equation~(\ref{eq:aod}), however, gives $\log N=13.46 \pm0.06$.  In what follows we will general take the lower column density for this system, but we note how our results would change if we adopted the higher value. We also detect two intervening \cfour systems at $z < 6.2$.  These detections are consistent with previous number density estimates in the literature, but does not provide significant additional constraints since this redshift range has been previously targeted by more extensive surveys. 

\begin{figure}
\includegraphics[width=\columnwidth]{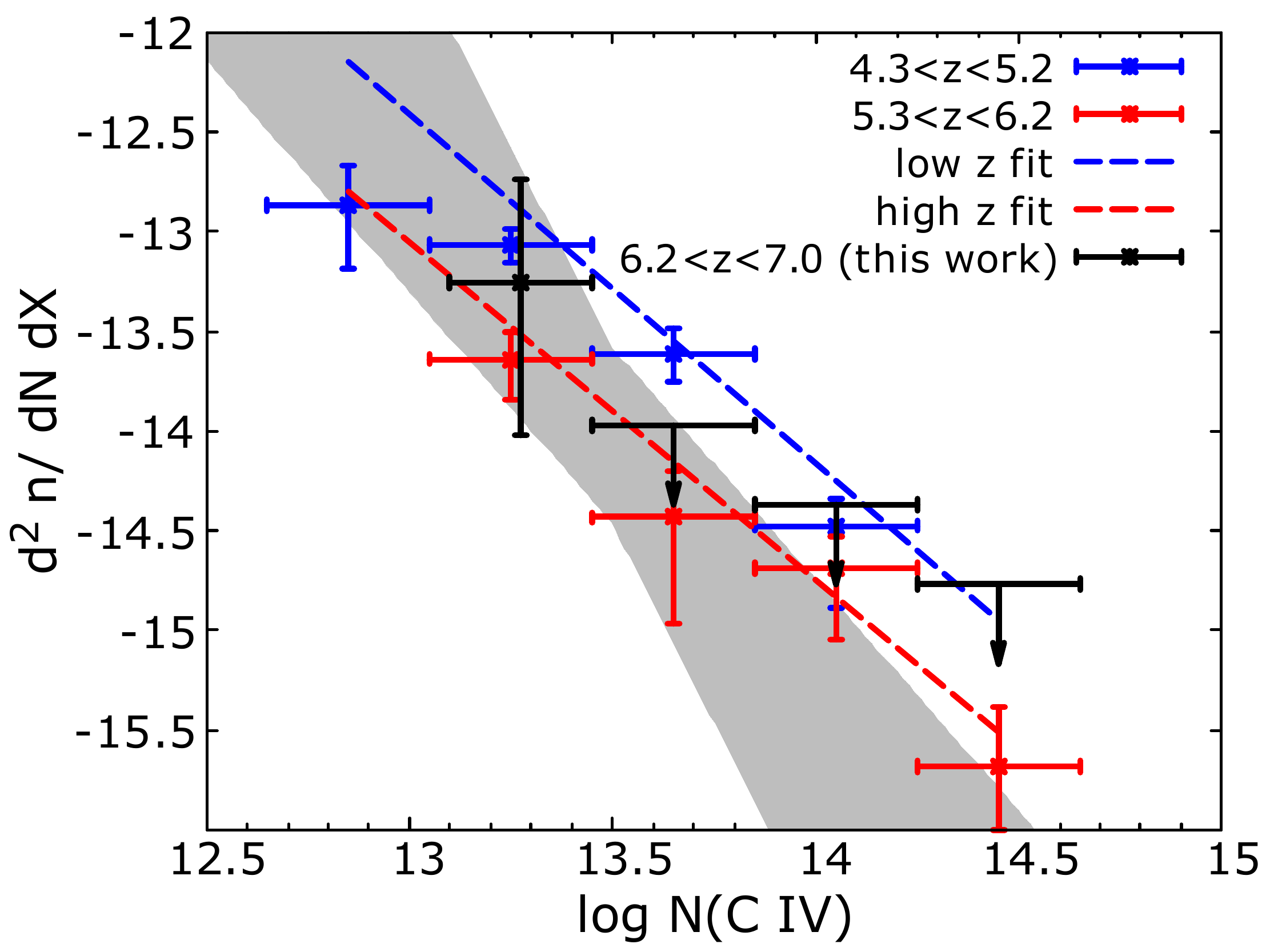}
\caption{Column density distribution of \cfour absorbers at $4.3<z<5.3$ (blue, D'Odorico et al 2013), $5.4<z<6.2$ (red, \citealt{Dodorico13} and J1120 line-of-sight) and $z>6.2$ (black, this work). Power-law fits are shown as dashed lines.  Given our pathlength and completeness, the detection of only a single system in our data, with column density $\log(N_{\text{C IV}} / \text{cm}^{-2})$ = 13.25, is consistent with the column density distribution at $z\sim5.5$, but marginally inconsistent with the $z\sim4.5$ distribution. The black upper limits correspond to 84 per cent single-sided Poisson uncertainties \citep{Gehrels86}. The gray contours show the $68\%$ confidence fit to the column density distribution (see text and Fig.~\ref{fig:cfour_likelihood}).
\label{fig:cfour_cddf}}
\end{figure}

\begin{figure}
\includegraphics[width=\columnwidth]{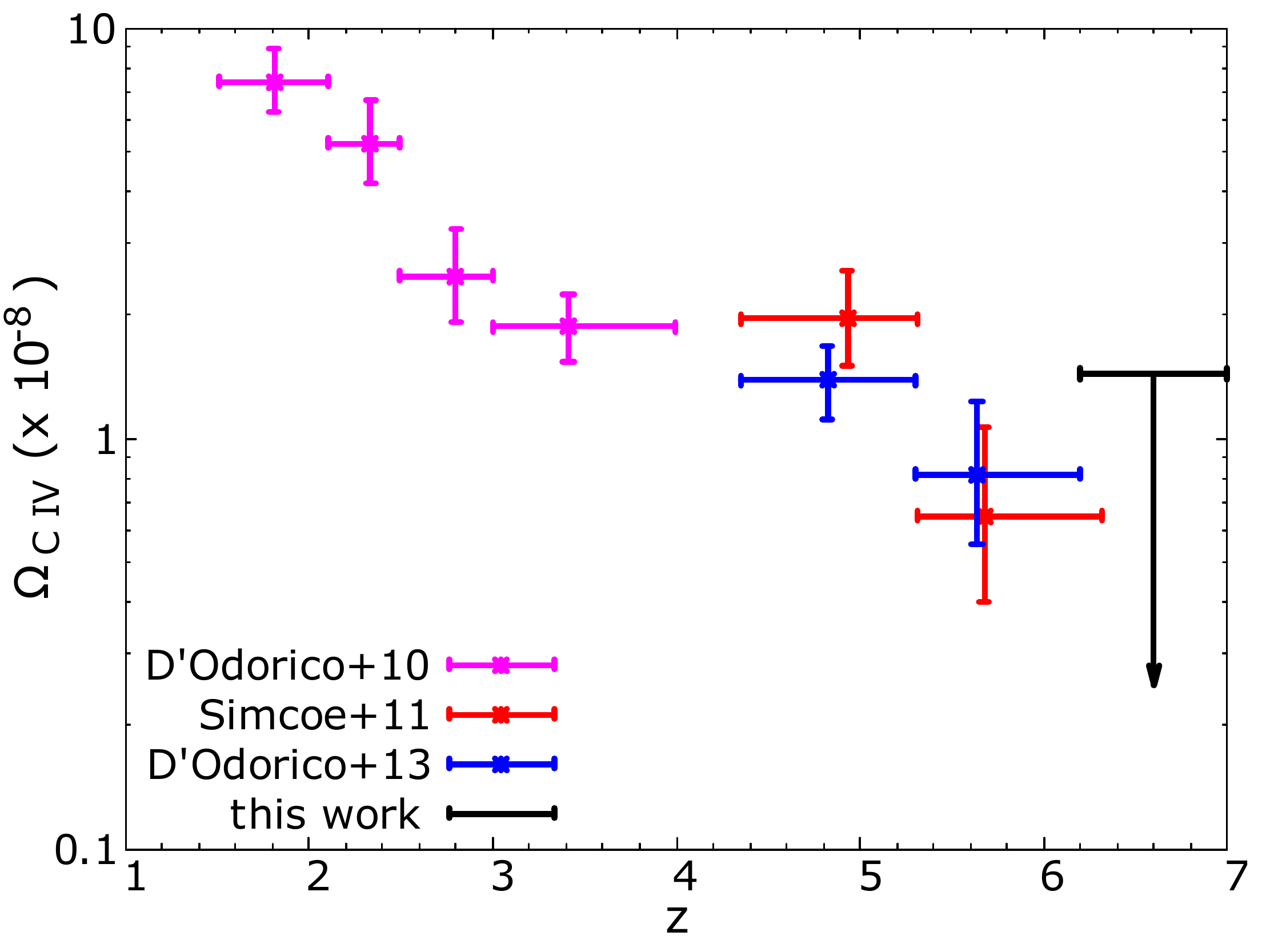}
\caption{Mass fraction of \cfour with redshift, including only strong absorption systems ($13.4 < \log(N_{\text{C IV}} / \text{cm}^{-2}) < 15.0$).  Our constraints are based on integrating over this column density range after using one detections with $\log(N_{\text{C IV}} / \text{cm}^{-2})$ = 13.46 (and the lack of other detections) to put constraints the underlying CDDF slope. }
\label{fig:omega_cfour}
\end{figure}

\begin{figure}
\vspace{-2.15em}
\hbox{\hspace{-0.8em}\includegraphics[width=1.11\columnwidth]{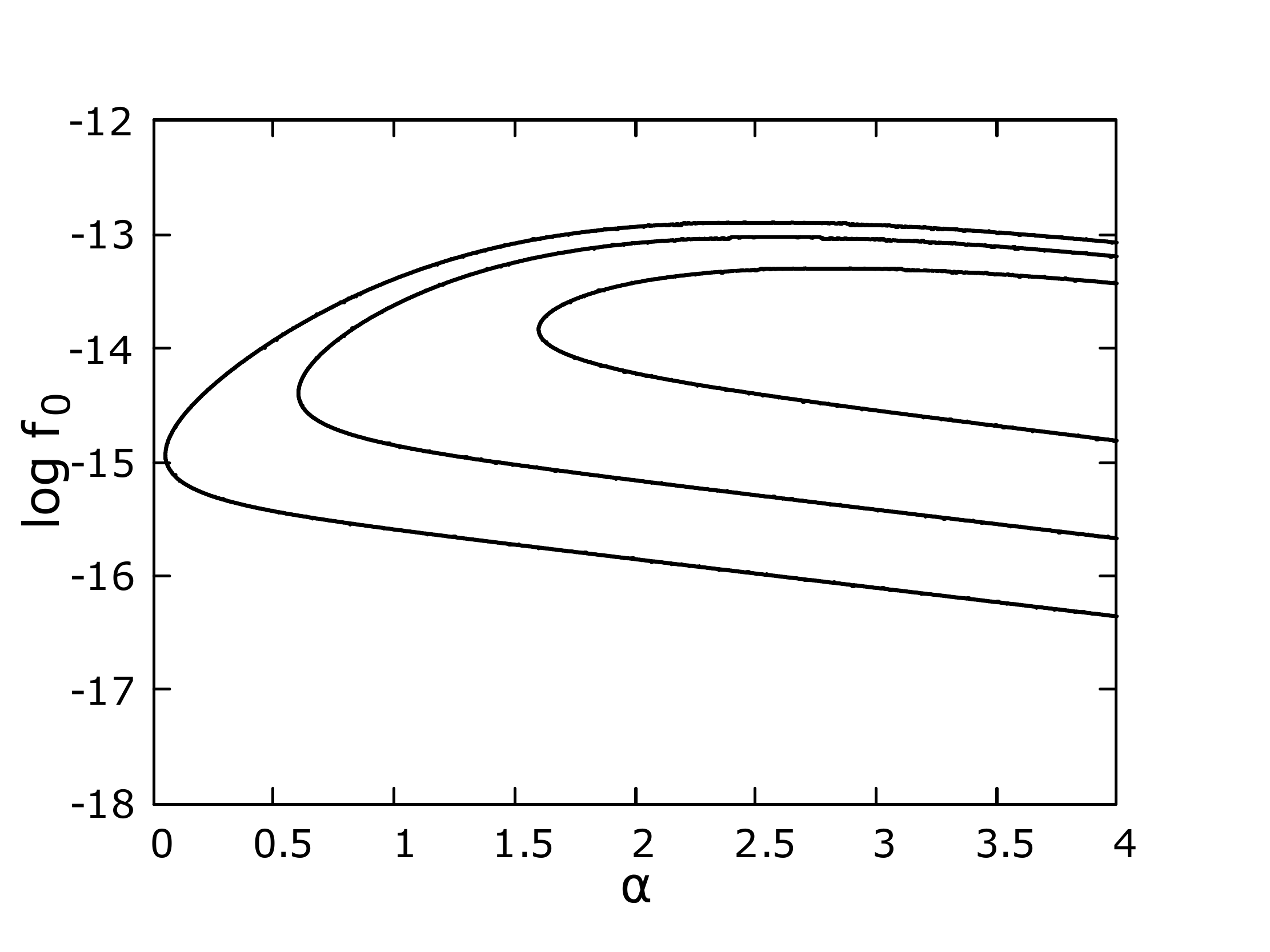}}
\caption{Posterior distribution of the \cfour distribution function parameters. Contours correspond to 68\%, 95\% and 99\% credible regions. The fit is made over $6.2<z<7.0$ and log $N > 13.1$, corresponding to our sensitivity threshold. Flat priors on $\alpha$ and $\log N_0$ are adopted (see text).}
\label{fig:cfour_likelihood}
\end{figure}

The constraints which can be obtained from a single detection are naturally weak. Nevertheless, we explore what constraints can be placed on the \cfour column density distribution and comoving mass density from our data.  Our binned column density results, corrected for completeness, are shown in Figure~\ref{fig:cfour_cddf}.  We also show lower-redshift data from \citet{Dodorico13}, along with power-law fits\footnote{The dashed line in Figure~\ref{fig:cfour_cddf} for $5.3 < z < 6.2$ is our own fit to the binned data from \citet{Dodorico13}.  We find a slope consistent with their value of $\alpha = 1.44$, but a lower best-fit normalization, $f(N_0) = 7.5 \times 10^{-15}~{\rm cm^{2}}$.} of the form
\begin{equation}
f(N) = f_0 \left( \frac{N}{N_0} \right)^{-\alpha} \, .
\label{eq:alpha}
\end{equation}
We use a value of $\log N_0$ = 13.5. We are roughly consistent with the column distribution of absorbers at $5.3 < z < 6.2$.  To obtain constraints on $\Omega_\text{C IV}$, we fit the column density distribution using a maximum-likelihood approach that jointly constrains the amplitude and slope of the column density distribution.  We define the likelihood function to be 
\begin{equation}
\mathcal{L}(f_0, \alpha) = P_{n}(n | f_0, \alpha) \times  \prod_i P_{i}\left( N_i | \alpha \right)  \, ,
\label{eq:cfour_likelihood}
\end{equation} 
where $P_{n}$ is the Poisson probability of observing the total number of systems in our sample, and $P_{i}$ is the probability of obtaining the the $i^{\rm th}$ column density.  All values are corrected for completeness.  For each value of $\alpha$ $P_{i}$ is taken from a distribution where $f_0$ has been chosen such that the expected mean number of systems is equal to the observed number.  Previous works have used a maximum likelihood approach to fit the slope, and then scaled the amplitude of the distribution assuming Poisson statistics (e.g., \citealt{Matejek12}).  The advantage of the present approach is that it properly accounts for the degeneracy between $\alpha$ and $f_0$, which is particularly important for small samples, and does so without binning the data.  We verified that our approach recovers appropriate best-fit values and credible intervals using mock samples.  We adopt a flat prior on $\alpha$ of $-4 \le \alpha \le 0$, which is equivalent to assuming that the distribution has not evolved dramatically from $z \sim 5.5$, for which \citealt{Dodorico13} find $\alpha \simeq 1.4$.  We fit over column densities $\log{N_\text{C IV}} \ge 13.1$, and use our completeness function for $b = 15~{\rm km\, s^{-1}}$.  The posterior distribution for $\alpha$ and $f_0$ is shown in Figure~\ref{fig:cfour_likelihood}. 
The marginalised constraints on individual parameters are $\log f_0 = -13.84_{-0.52}^{+0.38}$ and $\alpha > 2.32$ at $68\%$ confidence (credible interval). 
We use equation~(\ref{eq:omega}) to convert these results into constrains on $\Omega_\text{C IV}$.  Integrating over $13.4 \le \log{N_\text{C IV}} \le 15.0$, we find $\log{\Omega_\text{C IV}} = -8.7^{+0.5}_{-1.5}$. Here, $\log{\Omega_\text{C IV}} = -8.7$ is the probability-weighted mean value; the value with the maximum probability is $\log{\Omega_\text{C IV}} = -9.4$.  
Repeating the analysis using $\log N_\text{C IV} = 13.46$ for the system at $z=6.515$, we find $\log{\Omega_\text{C IV}} = -8.5^{+0.7}_{-1.4}$. The value at the peak probability is $\log{\Omega_\text{C IV}} = -8.9$.  
Adopting the higher column density naturally produces a higher $\Omega_\text{C IV}$, although the upper and lower limits increase only by a factor of two. Since we are only considering a finite range in $\alpha$, only the upper bound on $\Omega_\text{C IV}$ should be considered reliable. We plot this value in Figure~\ref{fig:omega_cfour} adopting the more conservative bound based on $\log N_\text{C IV, 6.515} = 13.46$ at $z = 6.515$.
Our results are consistent with a continuing decline in $\Omega_\text{C IV}$ with redshift (Figure~\ref{fig:omega_cfour}), albeit with large errors.  The implications of this result are discussed briefly in the next section. We note that models with increasingly negative $\alpha$ become indistinguishable; this is a consequence of the intrinsic degeneracy arising from fitting our data with a power law, and is reflected in the contours in Figure~\ref{fig:cfour_likelihood}.

\subsection{C II}\label{subsec:cii}

 Our sole \ctwo detection occurs at $z_\text{abs} = 6.4067$, and is identified via coincidence with \magtwo absorption.  The column density is $\log N_\text{C II} = 13.4 \pm 0.4$, where our completeness is $\sim$90 per cent.  \citealt{Becker11} find an incidence rate of $\text{d}n/\text{d}X \approx 0.25$ at $5.3 < z < 6.4$ using data of comparable quality to our spectrum of J1120 (although the system here is at the lower end of the range of $N_\text{C II}$ in that study).\footnote{We note that, unlike absorption doublets, singlet species such as C~{\scriptsize II} cannot be identified on their own.  Without the Lyman-$\alpha$ forest to flag potential low-ionisation absorbers via their H~{\scriptsize I} absorption, these ions must be identified via coincidence with other metal lines.  In this sense, we caution that the detection method for C~{\scriptsize II} is not consistent across surveys. \citealt{Becker11} lacked the near-infrared coverage to detect Mg~{\scriptsize II}, and searched for C~{\scriptsize II} based on coincidence with Si~{\scriptsize II} and O~{\scriptsize I}, while our C~{\scriptsize II} system was detected concurrently with Mg~{\scriptsize II}.  Care may therefore need to be taken when evaluating trends in C~{\scriptsize II} between different studies.}  For a non-evolving population we would expect to detect $\sim$1 system over our C~{\scriptsize II} pathlength of $\Delta X = 3.9$.  Our data thus presents tentative evidence that the number density of low-ionisation systems remains roughly constant\footnote{Although see \citealt{Becker11}, who find that the incidence rate of low-ionization systems may increase at $z \gtrsim 5.7$.} over $5 < z < 7$, which is in turn similar to the number of low-ionisation systems traced by neutral hydrogen absorbers with column densities $\log{N_\text{H I}} \ge 19.0$ (damped Lyman-$\alpha$ (DLA) and sub-DLA absorbers) over $3 < z < 5$.
  
Although the statistical claims that can be made from a single line-of-sight are naturally limited, our data point to an evolution in carbon over $6 \lesssim z \lesssim 7$ where the mass density of highly ionised \cfour declines with redshift while the number density of low-ionisation absorbers traced by \ctwo remains roughly constant.  This is broadly consistent with recent numerical models of chemical enrichment by star formation-driven galactic winds where the metals occupy increasingly higher densities and are exposed to an increasingly softer, spatially fluctuating ionizing background towards higher redshift (e.g., \citealt{Oppenheimer09}; \citealt{Finlator15, Finlator16}).  Significant numerical challenges remain in modeling these absorbers (e.g., \citealt{Keating16}; \citealt{Chen16}); however, the data presented here should provide additional leverage for testing numerical models in terms of their redshift evolution.

\subsection{Mg II}\label{sec:mgii}

\begin{figure}
\includegraphics[width=\columnwidth]{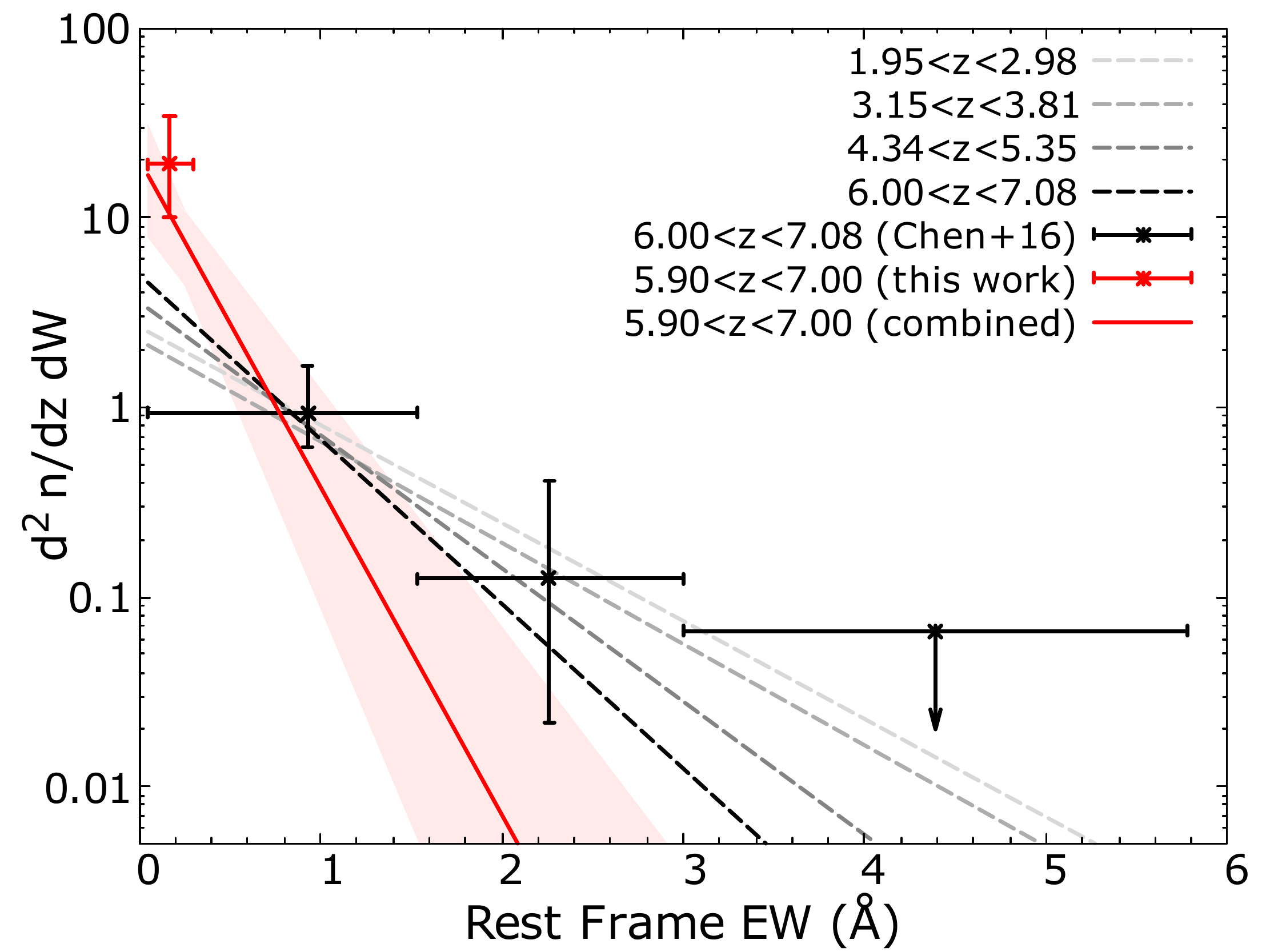}
\caption{Equivalent width distribution of Mg~{\small II} absorbers.  Data points are from \citet{Chen16} and this work.  Error bars assume Poisson statistics.  The dashed lines show the fits to the distribution at different redshifts from \citet{Chen16}.  The solid line is our fit to the distribution over $5.9 < z < 7.0$ using the combined datasets.  The shaded region is the 68 per cent credible region in the fit.  As discussed in the text, a single power law may not provide a sufficient description of the equivalent width distribution over the full range in $W$.}
\label{fig:mg2_distribution}
\end{figure}

Our deep X-Shooter spectrum is the first to be highly sensitive to very weak ($W < 0.3$) \magtwo systems  out to $z = 7$.
We searched for lines over $2.5<z<7.0$ with gaps around $ 3.9<z<4.1 $ and $ 5.6<z<5.9$ 
due to telluric absorption (see Fig. 3). We detect five \magtwo systems at $z>5.5$ (four at $z > 5.9$), all of which have $W < 0.5$ \AA, and three of which show additional low-ionisation ions (see Table~\ref{tab:summary}).  

\begin{figure}
\vspace{-2.15em}
\hbox{\hspace{-2.0em}\includegraphics[width=1.15\columnwidth]{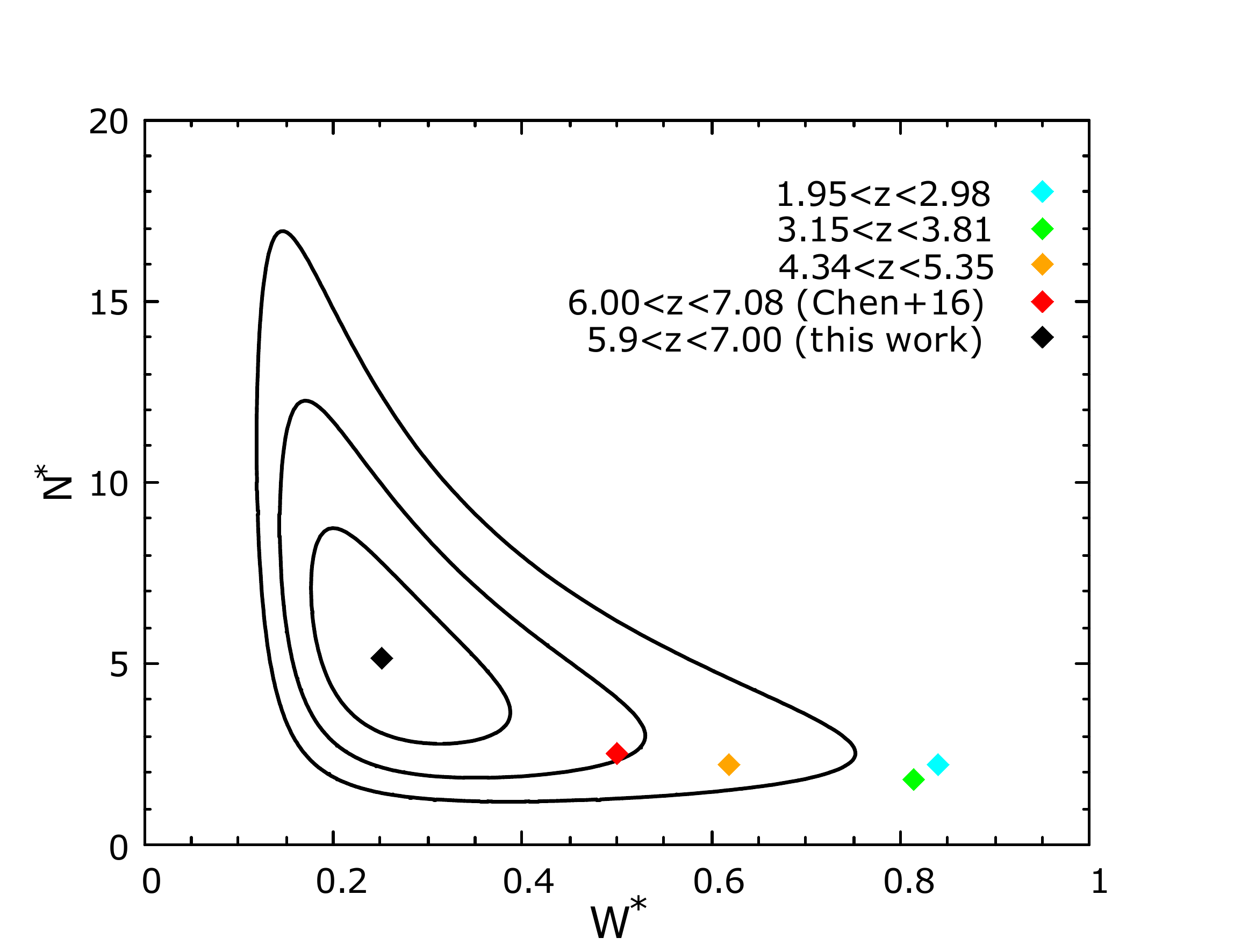}}
\caption{Posterior distribution of the \magtwo distribution function parameters. The best fitting parameters are indicated by a black diamond, with black contours corresponding to 68\%, 95\% and 99\% highest posterior density regions. The fit is made over $5.9<z<7.0$ (see text). Best fit parameters obtained by \citet{Chen16} at $\overline{z}=2.52, 3.46, 4.80, 6.28$ are shown as coloured diamonds. Our measurement is in 2$\sigma$ tension with previous work at the same redshift. }
\label{fig:mg2_likelihood}
\end{figure}

The equivalent width distribution of \magtwo absorbers has been shown (e.g., \citealt{Nestor05}) to be well described
 by an exponential function of the form
\begin{equation}
\frac{\text{d}^2 n}{\text{d}z\, \text{d}W} = \frac{N^*}{W^*} e^{-W/W^*} \, ,
\label{eq:d2ndzdW}
\end{equation}
at least for $W > 0.3$~\AA, a point we will return to below.  The scale factor $W^*$ peaks at $z\sim2.5$. Using the best-fit parameters from the highest redshift bin of \citet{Chen16} ($W^* = 0.50$~\AA, $N^* = 2.51$ over $6.00<z<7.08$; see Figure~\ref{fig:mg2_distribution}), the expected number of systems along the J1120 line-of-sight with $W>1$\ \AA \ over $5.9 < z < 7.0$ is $\sim$0.4, consistent with our non-detection of strong systems.\footnote{\citet{Chen16} also found no strong \magtwo systems towards J1120.}
For the same fit, however, we would expect to detect only $\sim$0.6 systems with $W < 0.3$~\AA, given our completeness, whereas we detect four (all with $W > 0.9$~\AA; see Table~\ref{tab:mgii_ew}).  Binned values of $\text{d}^2n/\text{d}z\text{d}W$ from this work and \citet{Chen16} are plotted in Figure~\ref{fig:mg2_distribution}.  Our binned value is estimated as $\text{d}^2n/\text{d}z\text{d}w \simeq \left(\sum{C_i^{-1}}\right)/\left(\Delta z \Delta W \right) = 19.1$, where $C_i$ is our completeness at the column density of the $i^{\rm th}$ absorber, $\Delta z = 1.1$, and $\Delta W = 0.25$, where our bin spans $0.05~{\rm \AA} < W < 0.3~{\rm \AA}$.  The error bars are Poisson.

\begin{table}
\caption{List of equivalent widths for \magtwo systems detected along the line-of-sight. Errors are measured from the error array.}
\centering
\begin{tabular}{l | c}
\hline
$z$ & $W/$\AA \\
\hline
6.40671 & 0.094 $\pm$ 0.022\\
6.21845 & 0.139 $\pm$ 0.029\\
6.1711 & 0.258 $\pm$ 0.057\\
5.9507 & 0.425 $\pm$ 0.060 \\
5.50793 & 0.455 $\pm$ 0.056\\
4.47260 & 0.276 $\pm$ 0.012\\
2.80961 & 0.246 $\pm$ 0.020\\
\hline
\end{tabular}
\label{tab:mgii_ew}
\end{table}

We re-evaluate the distribution of \magtwo systems at $z > 5.9$ by combining our data with those of 
\citet{Chen16} excluding their J1120 line-of-sight. The \citet{Chen16} sample is less sensitive but contains more lines of sight at $z > 6$, and therefore better constrains the the high-$W$ end of the distribution.  We use a maximum-likelihood, full Bayesian approach to constrain $W^*$ and $N^*$ similar to the one described in Section~\ref{sec:civ}.  Here, the combined likelihood function is given by
\begin{equation}
\mathcal{L}(W^*,N^*) = P_{n}(n | W^*, N^*) \times  \prod_i P_{i}\left( W_i | W^* \right)  \, ,
\label{eq:mg2_likelihood}
\end{equation}
where $P_{n}$ is the Poisson probability of detecting the total number of lines in our sample, and $P_i$ is the probability of obtaining the $i^{\rm th}$ equivalent width.  We use a redshift path-weighted mean completeness function that combines our J1120 data with the remainder of the sightlines from \citet{Chen16}.  The four \magtwo systems in our sample are combined with seven from \citet{Chen16} for a total of $n = 11$.  The fit is then performed over $0.05~{\rm \AA} < W < 5.0~{\rm \AA}$.

Our two-dimensional posterior is shown in Figure~\ref{fig:mg2_likelihood}.
 We find best-fitting values of  $W^* = 0.25_{-0.06}^{+0.09}$~\AA, $N^* = 5.128_{-1.78}^{+1.75}$, where the errors are 68 per cent marginalized credible regions.
The combined best-fit values of \citet{Chen16}, which used only stronger systems, are excluded at the $\sim$93 per cent level (Figure~\ref{fig:mg2_likelihood}).  It is not clear, moreover, that a single exponential provides a good fit over the full range of equivalent width.  With our best-fit parameters, for example, we would expect to detect $\sim$1.5 systems with $W < 0.3$~\AA, given our completeness, whereas we detect four, which would have a $\sim$5 per cent probability of occurring by chance for purely Poisson statistics.  These tensions may reflect a steepening of the equivalent width distribution at low equivalent widths.  Indeed,
\citet{Nestor05}, using observations by \citet{Churchill99}, first pointed out that the equivalent width distribution of \magtwo is more complicated than a single power law.  They fit a double exponential over $0.4<z<1.4$, finding an upturn in the number density of systems below $W=0.3$\AA.  More recently, \citet{Mathes17} used a Schechter function to fit the equivalent width distribution at $z < 2.6$, finding an exponential cut-off near $W \sim 2$~\AA.

\begin{figure}
\includegraphics[width=\columnwidth]{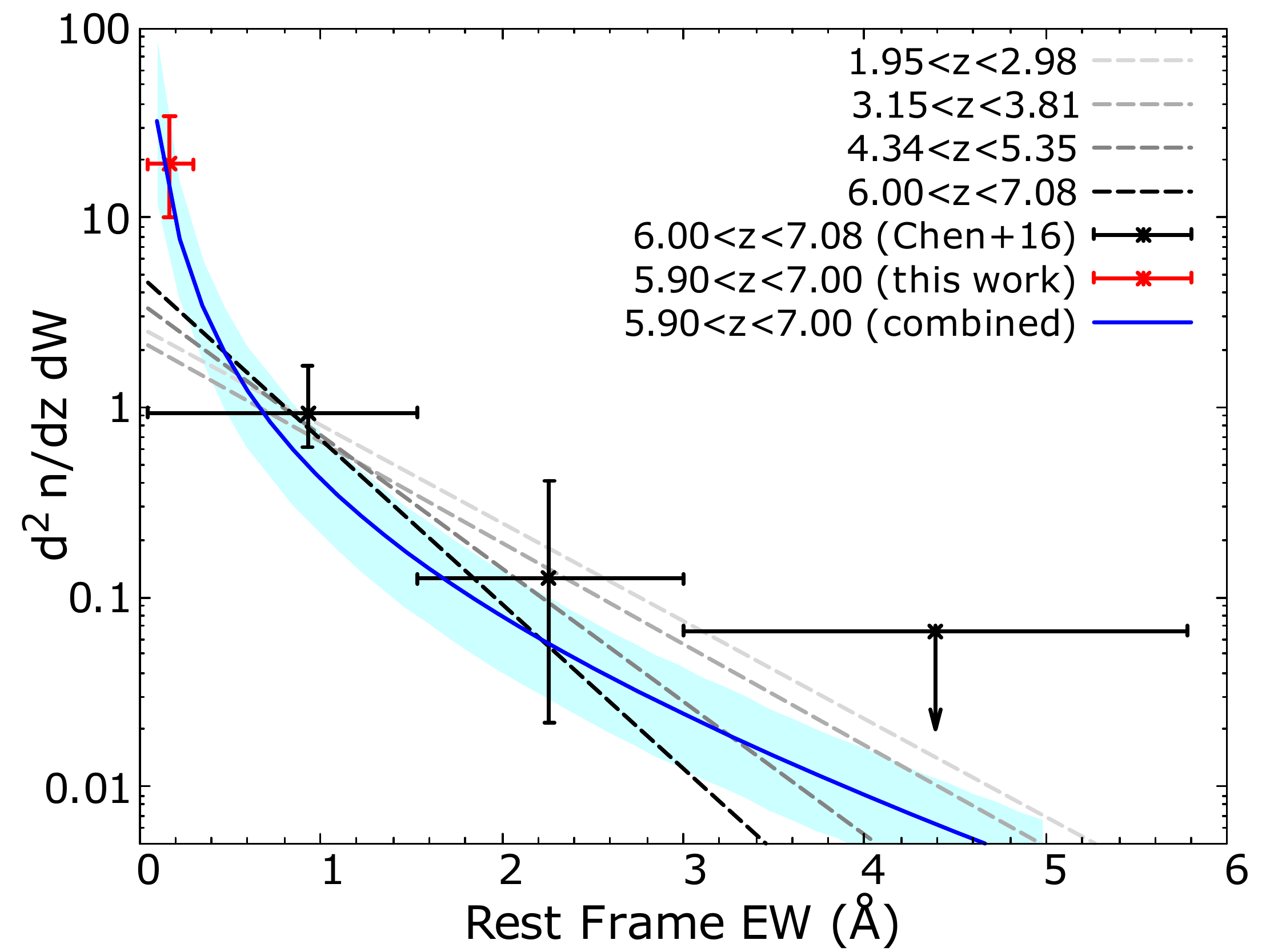}
\caption{Same as Figure~\ref{fig:mg2_distribution}, but fit with a Schechter function following Equation~\ref{eq:sch} where the turnover point is held fixed at $W^* = 2$\AA.}
\label{fig:sch}
\end{figure}

Following \citet{Kacprzak11} and \citet{Mathes17}, we to fit the combined sample of \citet{Chen16} and this line of sight with a Schechter function of the form:

\begin{equation}
\frac{\text{d}^2 n}{\text{d}z\, \text{d}W} = \left( \frac{\Phi^*}{W^*} \right) \left( \frac{W}{W^*}\right)^\alpha e^{-W/W^*}\label{eq:sch}
\end{equation}
which now depends on 3 parameters $\{W^*,\alpha,\Phi^*\}$. Unsurprisingly, the fit is highly unconstrained when all three parameters
are allowed to vary.
We nevertheless can investigate whether there is evidence of evolution compared to low redshift
 by fixing $W^* = 2$\AA, in agreement with the results of \citet{Mathes17} over the range $0.14<z<2.64$.
This yields best-fit values of $\Phi^* (\text{d}N / \text{d}z) = 0.43_{-0.21}^{+0.32}$, 
in agreement with lower redshift values\footnote{In the convention of \citet{Mathes17}: $\Phi^* (\text{d}N / \text{d}X) = 0.086_{-0.043}^{+0.064}$}; and $\alpha = -1.69 \pm 0.32$, in $\sim 2 \sigma$ tension with $z < 2.6$ results. We plot this fit in Figure~\ref{fig:sch}.

While the distribution function of \magtwo has not been probed in the weak regime at intermediate redshifts ($2.5 \lesssim z \lesssim 6$), the apparently high number of weak \magtwo systems we detect plausibly reflects complexity in the shape of the equivalent width distribution similar to what is seen at lower redshifts ($z \lesssim 2.5$).  In terms of their physical properties, weak \magtwo systems may not have the same origin at all redshifts.  Even so, it is possible that these weak systems at $z \sim 6$-7 are associated with accreting gas and/or the cooling remnants of previous metal-enriched outflows, as has been suggested for weak systems at $z \lesssim 2.5$ (see discussion in \citealt{Mathes17}).

\subsection{Associated Absorbers} \label{sec:associated}

In addition to intervening absorbers along the line-of-sight to J1120, we analysed 
absorbers close to the redshift of the QSO.
We find three such systems, located at $-2530$, $-1100$ and $-920$ km $\text{s}^{-1}$ blueward of the systemic redshift. The strongest system, at $-1100$ km $\text{s}^{-1}$, was identified in \cfour and \nfive in the discovery spectrum by \citet{Mortlock11}.  The two remaining systems, as well as the \sfour in the strongest system, are newly identified here.

\noindent
The system at $z=7.01652$ (2530 km$\text{s}^{-1}$ from the QSO's systemic redshift) contains weak \cfour absorption and can be seen in Figure~\ref{fig:last}. The two highest-redshift systems, at $z=7.05540$ and $z=7.06001$, consist of C~{\small IV} and N~{\small V}, as well as Si~{\small IV} absorption in the former.\footnote{The fact that these systems occur blueward of the QSO redshift yet the \cfour\ lines fall on the red side of the \cfour\ emission line (Figure~\ref{fig:spectrum}) reflects the extreme blueshift of this object's \cfour\ emission line, noted by \citet{Mortlock11}.}

The associated systems at $z \simeq 7.06$ display unusual absorption profiles, as the apparent optical depths of the \cfour and \nfive doublets are in ratios $\tau_{1548} / \tau_{1550} = 1.32$,  $\tau_{1238} / \tau_{1242} = 1.21$, respectively, for the $z=7.060$ system and 
$\tau_{1548} / \tau_{1550} = 1.05$,  $\tau_{1238} / \tau_{1242} = 1.15$ for the $z=7.055$ system.
While saturation can drive the equivalent width ratios below the canonical value of 2:1 expected for optically thin lines, for the $z=7.055$ system the large residual flux ($\sim40$ per cent) makes it unlikely that saturation is the only effect (see Figures 11 and 12). 
Two plausible explanations of the peculiar ratio of the doublets are (i) the intervening system contains a column density of \cfour and \nfive sufficient for saturated absorption, but covers only a fraction of the continuum source -- {\textit{partial covering}}, or (ii) the absorption feature is composed of multiple unresolved components, each individually saturated and located close enough in velocity space to appear blended.

The {\tt{vpfit}} program was adjusted to test the relative goodness of fit provided by these two possibilities. To test partial covering, an additional variable representing the covering fraction of the continuum was added to the fit. 
The C~{\small IV}, \nfive and Si~{\small IV} lines were fit simultaneously, with the covering fraction, redshift and Doppler parameter of each component constrained to be the same for all ions.
Multiple narrow absorbers, on the other hand, were fit by modifying the initial conditions of the fit to contain two narrow absorbers for each absorption line.  The starting values of the Doppler parameters
were initially forced to be $b<7$ km $\text{s}^{-1}$. 
After letting the fit converge this condition was relaxed and the fit was re-run.   Component redshifts and Doppler parameters were once again tied between ions.
In both these fitting techniques, we introduced an extra `slope' parameter over each fitting window to allow for adjustments to the continuum normalisation.

\begin{figure}
\includegraphics[width=0.97\columnwidth]{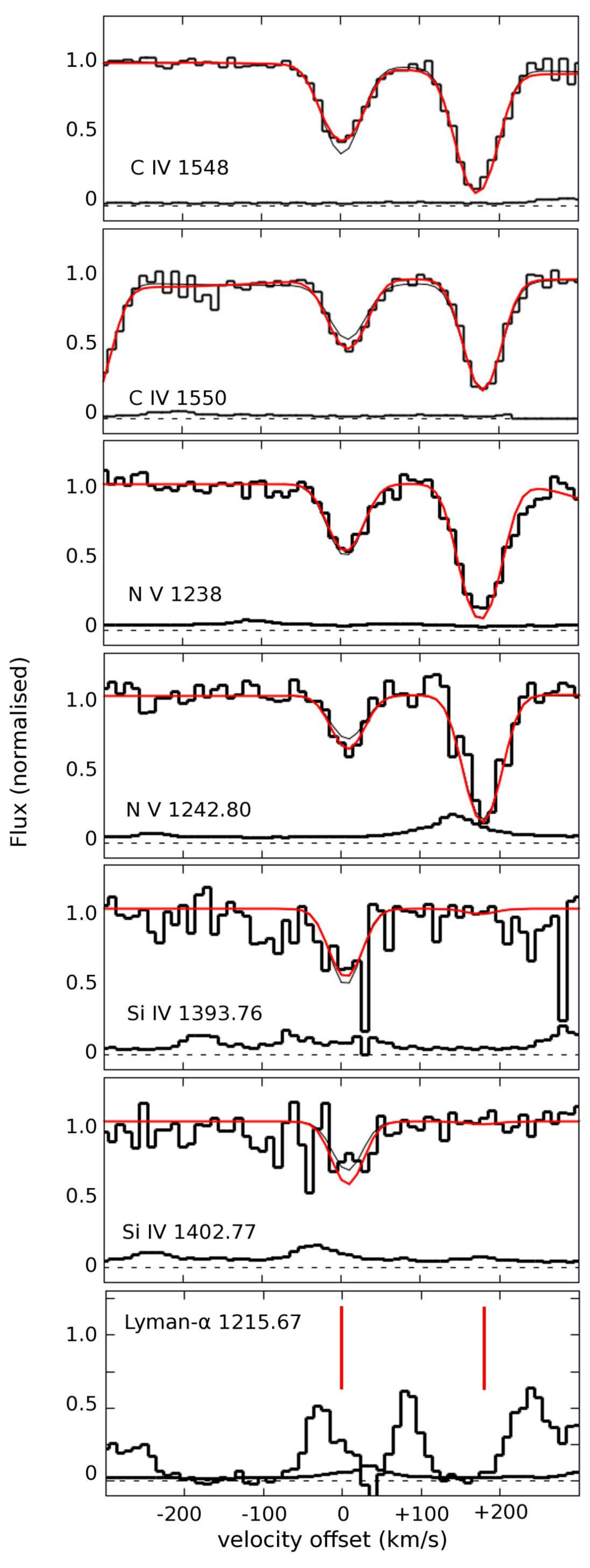}
\caption{Associated absorption systems at $z=7.055$ and $z=7.060$.  The plot is centered at $z=7.055$.  Thick continuous lines show single-component fits using partial covering (Table~\ref{tab:partial}).  Thin lines show fits without partial covering (Table~\ref{tab:naive}).  The locations of \lal for these components are indicated with tick marks in the bottom panel.}
\label{fig:single}
\end{figure}

Narrow absorbers and partial covering provide comparably good fits to the data, both improving upon naive single-component fits (Table~\ref{tab:naive}).  The best-fit partial covering for the $z=7.055$ system is $42 \pm 2$ per cent (see Table~\ref{tab:partial}), with a $\chi^2$ per degree of freedom integrated over all components of 2.641.
In both this fit and the alternative, the $\chi^2$ is driven primarily by the Si~{\small IV} doublet, which may suggest that the noise in the Si~{\small IV} line exceeds the estimate in the error array (see Figure~\ref{fig:single}).  Omitting \sfour, the reduced $\chi^2$ is 1.736.
Multiple unresolved absorbers provide a similarly good fit to the data, but the column density of one of the components is highly unconstrained for all ions (Table~\ref{tab:narrow}). The $\chi^2$ per degree of freedom of the fit is 2.656, dominated again by the Si~{\small IV} doublet, or 1.772 when omitting \sfour.

\begin{figure}
\includegraphics[width=0.97\columnwidth]{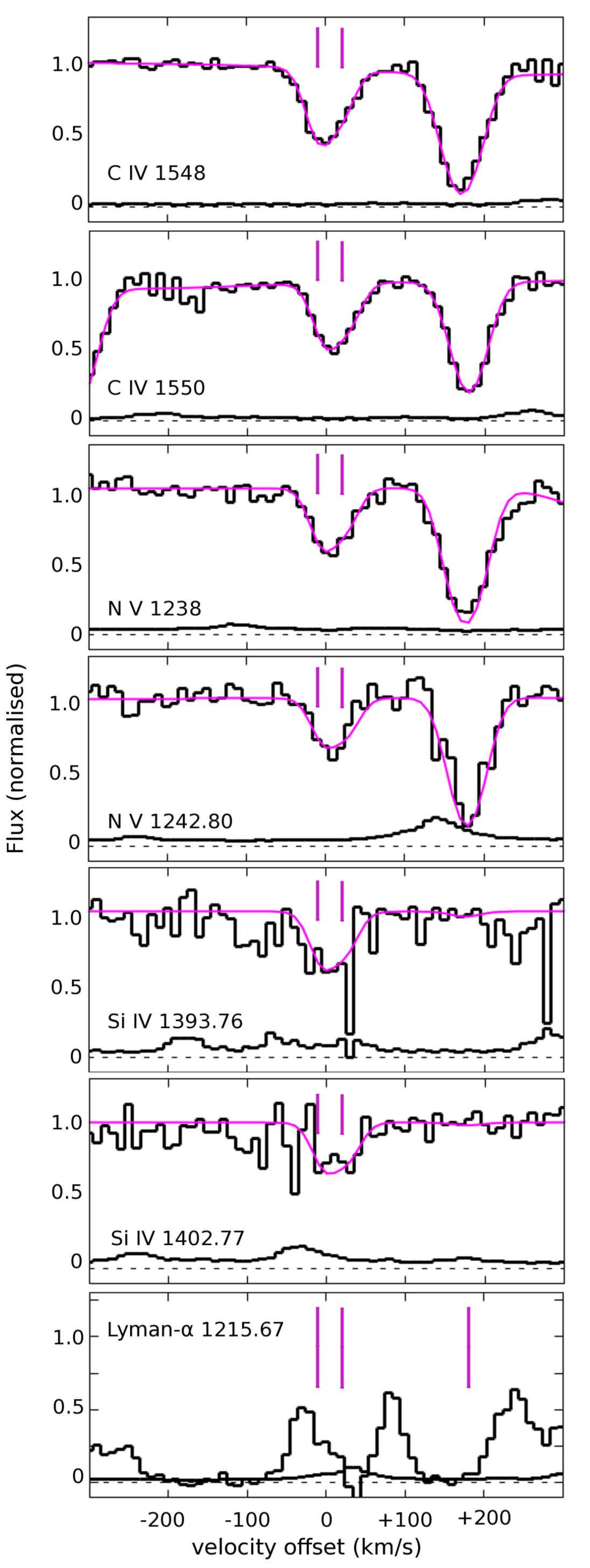}
\caption{Same as Figure~\ref{fig:single}, but here the thick continuous lines show fits using multiple, unresolved components for the system at $z=7.055$ (Table~\ref{tab:narrow}).  The velocities of the components are indicated with vertical tick marks.}
\label{fig:assoc_multi}
\end{figure}

\clearpage

\begin{table*}
\caption{Best-fit parameters for the fits to associated systems using single components and no partial covering.}
\begin{tabular}{l | c | c | c | c}
\hline
$z_\text{abs}$ & log $N_{\text{C IV}}/\text{cm}^{-2}$ & log $N_{\text{N V}}/\text{cm}^{-2}$ & log $N_{\text{Si II}}/\text{cm}^{-2}$ & $b$ \\
\hline
 $7.05541 \pm 0.00002$& 13.88 $\pm$ 0.02 & 13.87 $\pm$ 0.03 & 13.34 $\pm$ 0.09 & $21.8 \pm 1.4$ \\
$7.060000 \pm 0.000013$& 14.44 $\pm$ 0.04& 14.7 $\pm$ 0.09&11.9 $\pm$ 0.5&$19.3 \pm 1.0$\\
\hline
 \multicolumn{5}{c}{$\chi^2 / N_\text{dof} = 2.950$} \\
\multicolumn{5}{c}{$\chi^2_{\text{no Si IV}} / N_\text{dof} = 2.115$} \\
\end{tabular}
\label{tab:naive}
\end{table*}

\begin{table*}
\caption{Best-fit parameters for the fits to associated systems using single components and allowing for partial coverage of the continuum.}
\begin{tabular}{l | c | c | c | c | c}
\hline
$z_\text{abs}$ & $\text{F}_\text{cover}$ & log $N_{\text{C IV}}/\text{cm}^{-2}$ & log $N_{\text{N V}}/\text{cm}^{-2}$ & log $N_{\text{Si II}}/\text{cm}^{-2}$ & $b$ \\
\hline
 $7.05541 \pm 0.00002$&  $42 \pm 2$ per cent & 14.6 $\pm$ 0.2 &  14.37 $\pm$ 0.11 & 14.2 $\pm$ 0.3 & $17.9 \pm 1.8$ \\
$7.060000 \pm 0.000013$& $10 \pm 5$ per cent&14.44 $\pm$ 0.04 &14.82 $\pm$ 0.12 & 11.9 $\pm$ 0.5 &$18.8 \pm 1.1$\\
\hline
  \multicolumn{6}{c}{$\chi^2 / N_\text{dof} = 2.641$} \\
  \multicolumn{6}{c}{$\chi^2_{\text{no Si IV}} / N_\text{dof} = 1.736$} \\
\end{tabular}
\label{tab:partial}
\end{table*}

\begin{table*}
\caption{Best-fit parameters for the fits to associated systems using single components and no partial covering.}
\begin{tabular}{l | c | c | c | c}
\hline
$z_\text{abs}$ & log $N_{\text{C IV}}/\text{cm}^{-2}$ & log $N_{\text{N V}}/\text{cm}^{-2}$ & log $N_{\text{Si II}}/\text{cm}^{-2}$ & $b$ \\
\hline
$7.05514 \pm 0.00004$ &14.2 $\pm$ 0.3& 13.94 $\pm$ 0.15&  13.6 $\pm$ 0.5 &$5.8 \pm 1.8$\\
$7.05596 \pm 0.00006$ &14.3 $\pm$ 1.4 & 14.1 $\pm$ 1.0& 14.5$\pm$ 1.6 &$ 2.7 \pm 1.6$\\
$7.060002 \pm 0.000013$ & 14.45 $\pm$ 0.04&14.84 $\pm$ 0.12 & 11.9 $\pm$ 0.5 & $18.7 \pm 1.1$ \\
 \hline
  & \multicolumn{3}{c}{$\chi^2 / N_\text{dof} = 2.656$} \\
  & \multicolumn{3}{c}{$\chi^2_{\text{no Si IV}} / N_\text{dof} = 1.772$} \\
\end{tabular}
\label{tab:narrow}
\end{table*}

The $\chi^2$ results do not allow one model to be definitely preferred over the other. 
The multiple component model, however, may be less viable on physical grounds.
Using $b = \sqrt{2kT/m}$, a $b$-parameter of 2.7 km$\text{s}^{-1}$ (4.3 km$\text{s}^{-1}$  at the 1$\sigma$ upper limit) would set an upper limit on the temperature of the \cfour gas of $T \lesssim 5000 \text{K}$ ($T \lesssim 12000 \text{K}$), which is potentially problematic
 if the gas is photo-ionised by the QSO. 
Partial covering in associated narrow QSO absorbers, on the other hand, is a well-documented phenomenon (\citealt{Misawa07};  \citealt{Wu10}; \citealt{Simon12}; \citealt{Dodorico04}).

The partial covering hypothesis, if correct, could have significant implications for the proximity zone of J1120.  As noted by \citet{Simcoe12}, \lal at the redshift of these metal absorbers is unsaturated -- our spectrum confirms this. Lack of saturation would normally indicate that the H~{\small I} column density is too low to be optically thick to ionizing photons ($\log N_\text{H I} < 17.2$).  If partial covering is a factor, however, then optically thick H~{\small I} may indeed be present, but suppressing only part of the QSO continuum.  This is more likely for the component at $z=7.055$, which contains Si~{\small IV} and is probably less highly ionized than the component at $z=7.060$.  Even a partial suppression of the ionizing continuum could contribute to the apparent shortness of the proximity zone noted by \citet{Mortlock11}. While it is difficult to know whether this scenario is correct for J1120, it may be of interest as data for further QSOs at $z > 7$ are obtained.

\section{Summary}\label{sec:summary}

We have used a deep (30h) X-Shooter spectrum of the $z = 7.084$ QSO \ulas to probe absorption by multiple metal species up to the highest redshifts to date. 
We find seven intervening systems in the range $5.5<z<7.0$  and three associated systems. The intervening systems span a wide range of ionic compositions and velocity profiles.
Our main results are: 

\begin{enumerate}

\setlength\itemsep{6pt}

\item We detect a single \cfour system at $z > 6.2$, which is a relatively weak absorber (log $\text{N}_{C IV}$ = 13.25 $\pm$ 0.06) at $z=6.51$.  Using a maximum likelihood method to set limits on the column density distribution, we demonstrate that the inferred comoving \cfour\ mass density at $z > 6.2$ is consistent with a continuous decline over $4<z<7$, though non-evolution from $z \sim 5.5$ cannot be ruled out.

\item We find one \ctwo absorber over $6.3 < z < 7.0$, consistent with the incidence rate of low-ionization absorbers at $z \sim 6$.  A decline in \cfour with redshift and a relatively flat evolution in \ctwo would be consistent with models that combine lower overall enrichment and a softer ionising background towards higher redshifts.

\item We identify four weak ($W < 0.3$~\AA) \magtwo systems, which exceeds predictions based on an extrapolation of a power law fit to the incidence rate of stronger systems at these redshifts \citep{Chen16}.  This is reminiscent of a similar enhancement in the number density of weak systems at $z < 2.5$ (e.g., \citealt{Nestor05}), which are potentially associated with inflows and/or cooling fragments of metal-enriched outflows (e.g., \citealt{Mathes17}).

\item We also investigate N~{\small V}, C~{\small IV}, and Si~{\small IV} systems associated with the QSO itself. One system located $\sim$$-1100$ km s$^{-1}$ blueward of the QSO shows peculiar absorption profiles in the \cfour and \nfive doublets in terms of the relative strengths of the doublet lines. Two explanations, partial covering of the continuum source and multiple unresolved components, are tested and found to explain this effect comparably well.  Multiple narrow components provide a reasonable fit; however, we argue that this scenario is physically unlikely as it would require photoionised gas within $ +1000 \text{km s}^{-1}$ of the QSO to have a temperature $T \lesssim 5000$ K ($T \lesssim 12000$ K using the upper 1$\sigma$ bound on $b$).
Alternatively, a single-component absorber with a covering fraction of $\sim$40 per cent would produce a similar line profile.
In this scenario, a partially covered hydrogen Lyman limit system could also be present even though \lal at the redshift of the metal absorber is not saturated.  Such a scenario could potentially help explain the apparent shortness of J1120's proximity zone.

\end{enumerate}

It is worth emphasizing that these results are based on only one line-of-sight.  We have often estimated uncertainties using Poisson statistics, which may under-estimate the scatter between lines of sight if metal absorbers are significantly clustered at these redshifts.  In addition to large-scale density variations, clustering due to fluctuations in the ionising background could also play a role.
The recent discovery at $z\sim5.5$ of a contiguous $\sim 110$ comoving Mpc trough of opaque \lal absorption by \citet{Becker15} illustrates the fact that ionisation conditions in diffuse gas can be correlated over large distances at these redshifts. 
Nevertheless, while the information which can be gained from a single line-of-sight is limited, it provides a glimpse into the circum-galactic media of some of the earliest galaxies, and therefore into key mechanisms governing galaxy formation.
Metal lines are powerful tools for studying the high redshift universe, and future studies 
should shed further light on the trends hinted at here.

\section*{Acknowledgements}\label{sec:ack}

The authors thank Stephen Chen for providing their $z > 6$ completeness function excluding the J1120 line-of-sight.  We also thank Steve Warren and Xiaohui Fan, and the anonymous referee, for helpful comments that substantially improved the manuscript.
Based on observations collected at the European Southern Observatory, Chile, programmes 286.A-5025(A), 089.A-0814(A), and 093.A-0707(A).
Support by the ERC Advanced Grant Emergence -- 32056 is gratefully acknowledged.
SEIB was supported by
a Graduate Studentship from the Science and
Technology Funding Coucil (STFC).  GDB was supported by the NST under grant AST-1615814.
BPV acknowledges funding through ERC grant Cosmic Dawn.

\bibliographystyle{apj} \bibliography{Metals_paper_resubmit}

\appendix
\section{Spectra of intervening systems}\label{sec:spectra}
Here we plot the X-Shooter spectrum of \ulas at the location of the detected absorbers. The range shown covers $\Delta v = \pm 400$ km $\text{s}^{-1}$; the pixel size is 10 km $\text{s}^{-1}$. 
Shaded regions highlight the detected lines.

\begin{figure}
\includegraphics[width=\columnwidth]{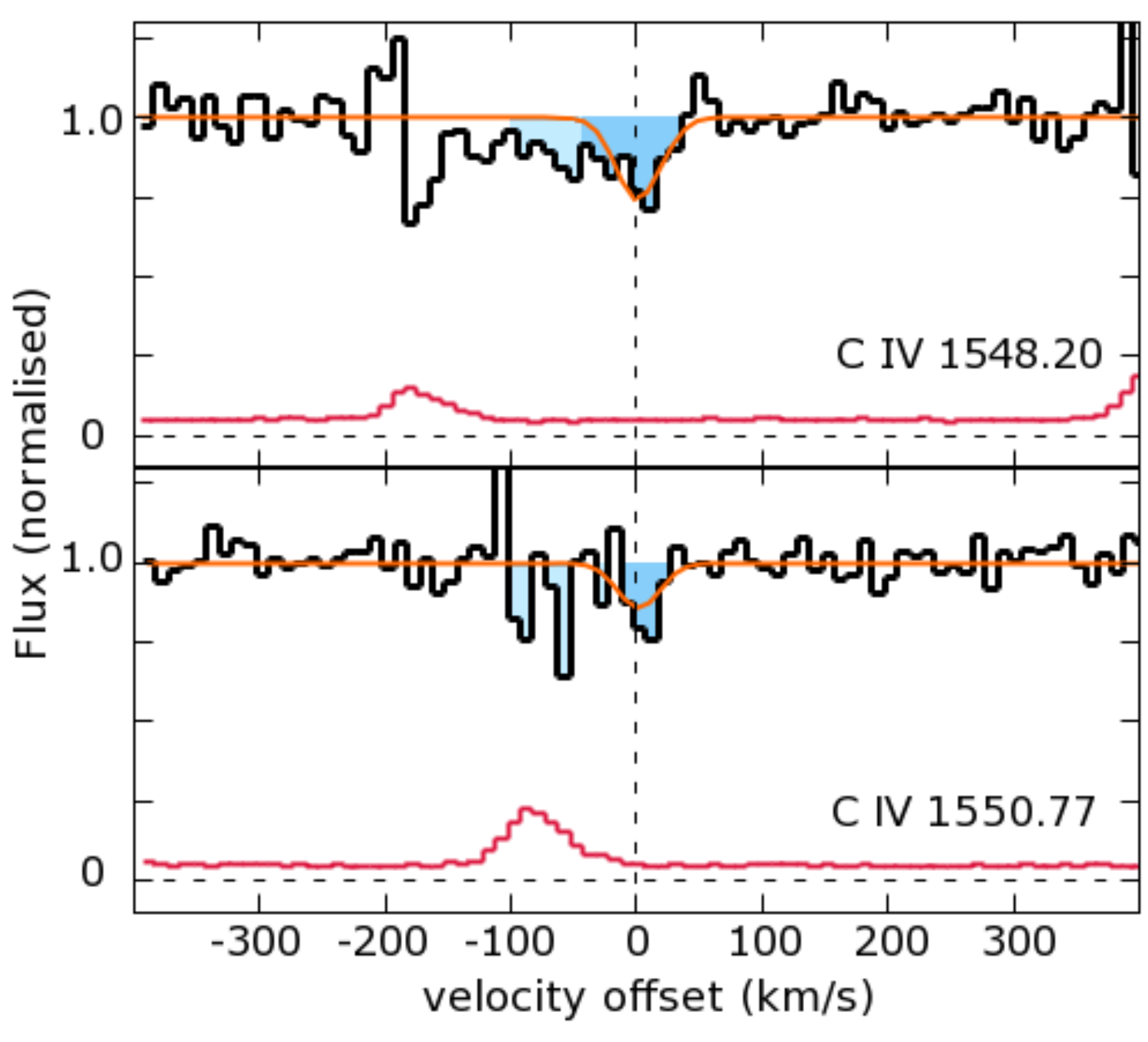}
\caption{Intervening system at $z=6.51511$.  The thick solid histogram shows the flux.  The thin histogram is the error array.  Ion identifications are printed at the bottom right of each panel. The orange line shows the best fit Voigt profile returned by {\tt{vpfit}} (see text). The blue wing of the $\lambda$1550.77 line is affected by skyline residuals.  The detection of absorption at $\Delta v < -20$ km s$^{-1}$ (shaded) is therefore tentative for this system.
\label{fig:cfour_z6p51}}
\end{figure}

\begin{figure}
\includegraphics[width=\columnwidth]{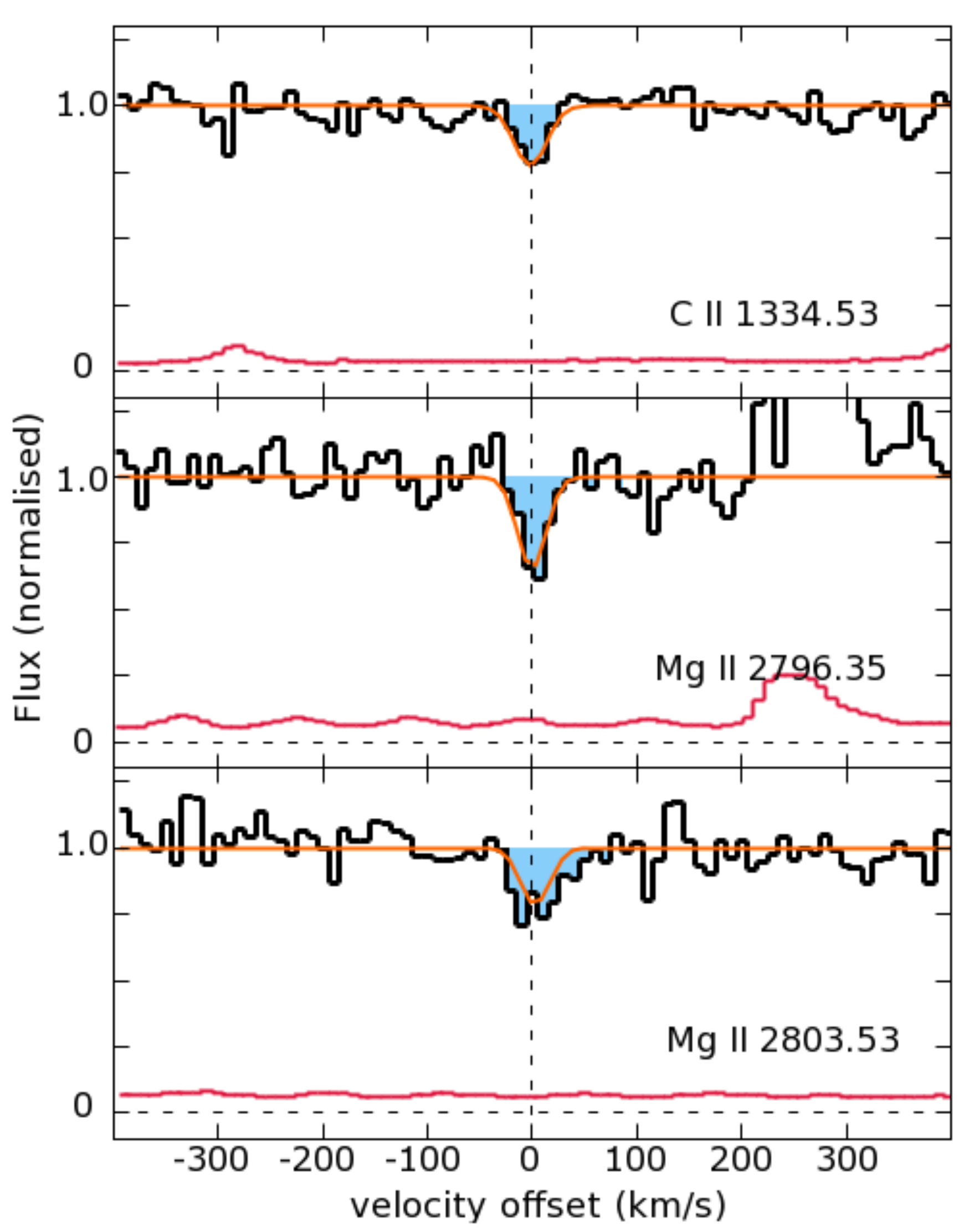}
\caption{Intervening system at $z=6.40671$.  Lines are as in Figure~\ref{fig:cfour_z6p51}. }
\end{figure}

\begin{figure}
\includegraphics[width=\columnwidth]{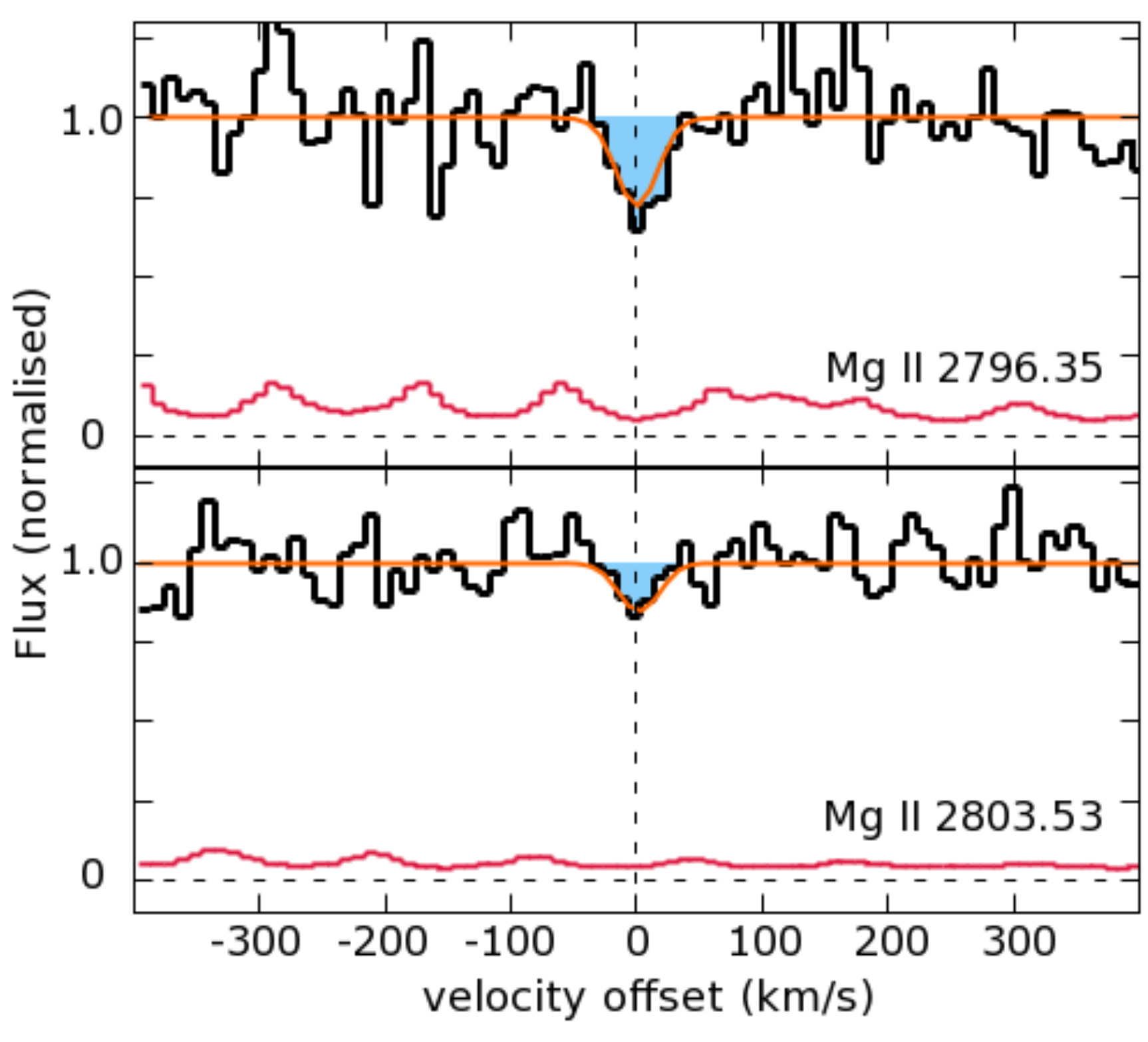}
\caption{Intervening system at $z=6.21845$.  Lines are as in Figure~\ref{fig:cfour_z6p51}.}
\end{figure}

\begin{figure}
\includegraphics[width=\columnwidth]{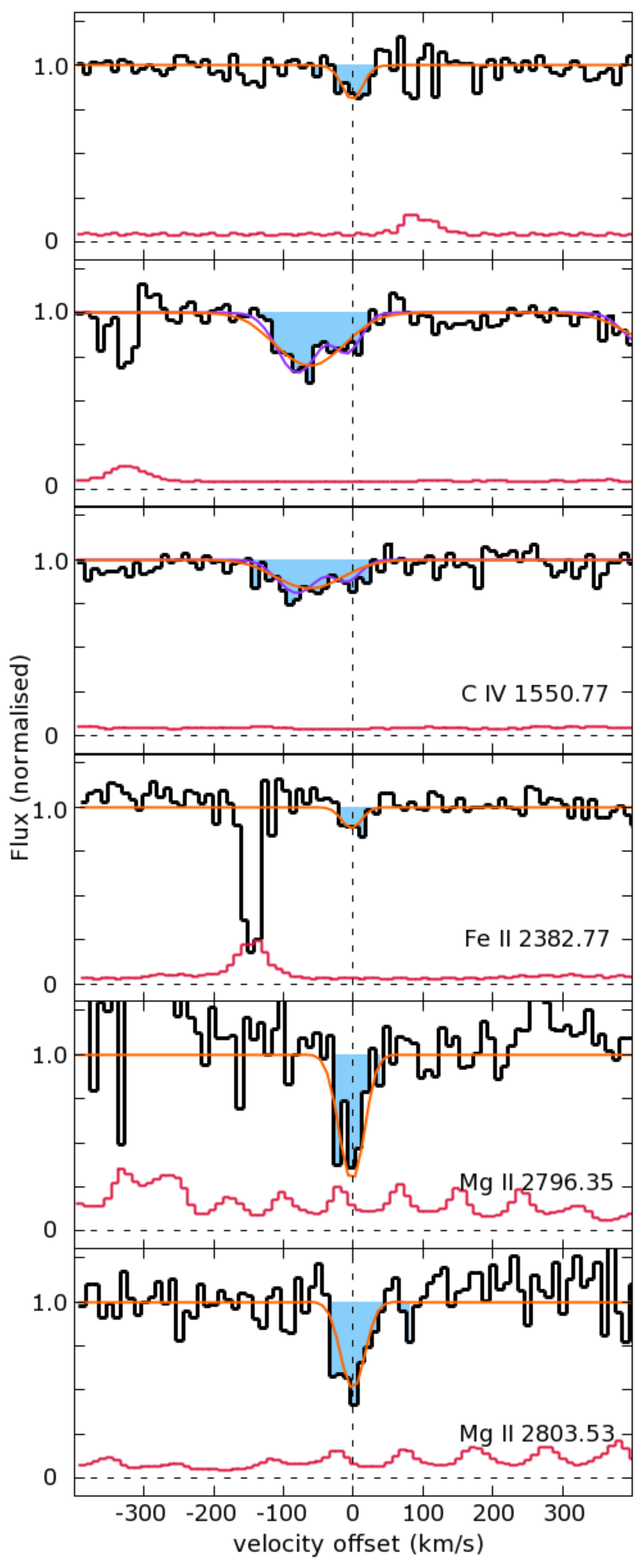}
\caption{Intervening system at $z=6.17110$.  Lines are as in Figure~\ref{fig:cfour_z6p51}. The thin purple lines present a possible two-component fit to the system, with best-fit parameters $\text{log} N_1 = 13.54 \pm 0.04$, $\text{log} N_2 = 13.18 \pm 0.09$ with $z_{1,2} = 6.1691, 6.1708$.}
\end{figure}

\begin{figure}
\includegraphics[width=\columnwidth]{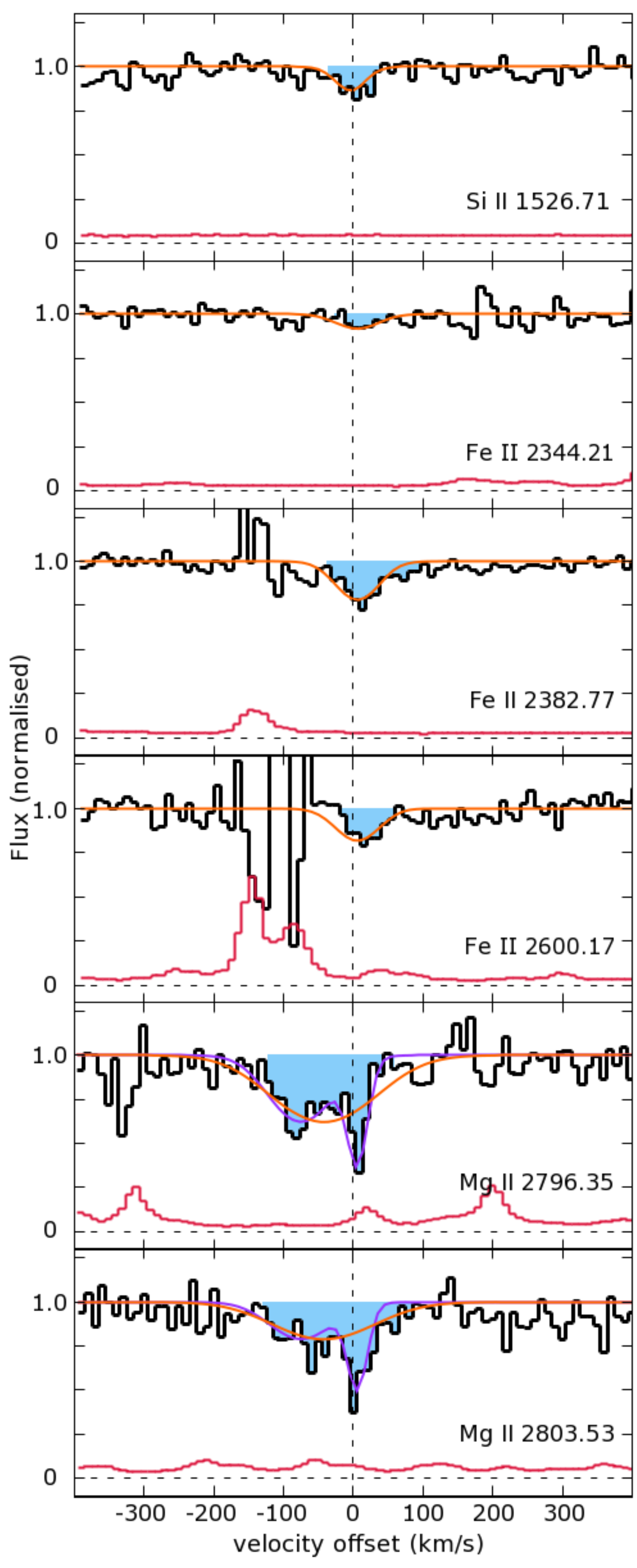}
\caption{Intervening system at $z=5.9507$.  Lines are as in Figure~\ref{fig:cfour_z6p51}. The thin purple lines present a possible two-component fit to the system, with best-fit parameters $\text{log} N_1 = 12.8 \pm 0.1, \text{log} N_2 = 12.9 \pm 0.1$ with $z_{1,2} = 5.9486, 5.9507$. }
\end{figure}

\begin{figure}
\includegraphics[width=\columnwidth]{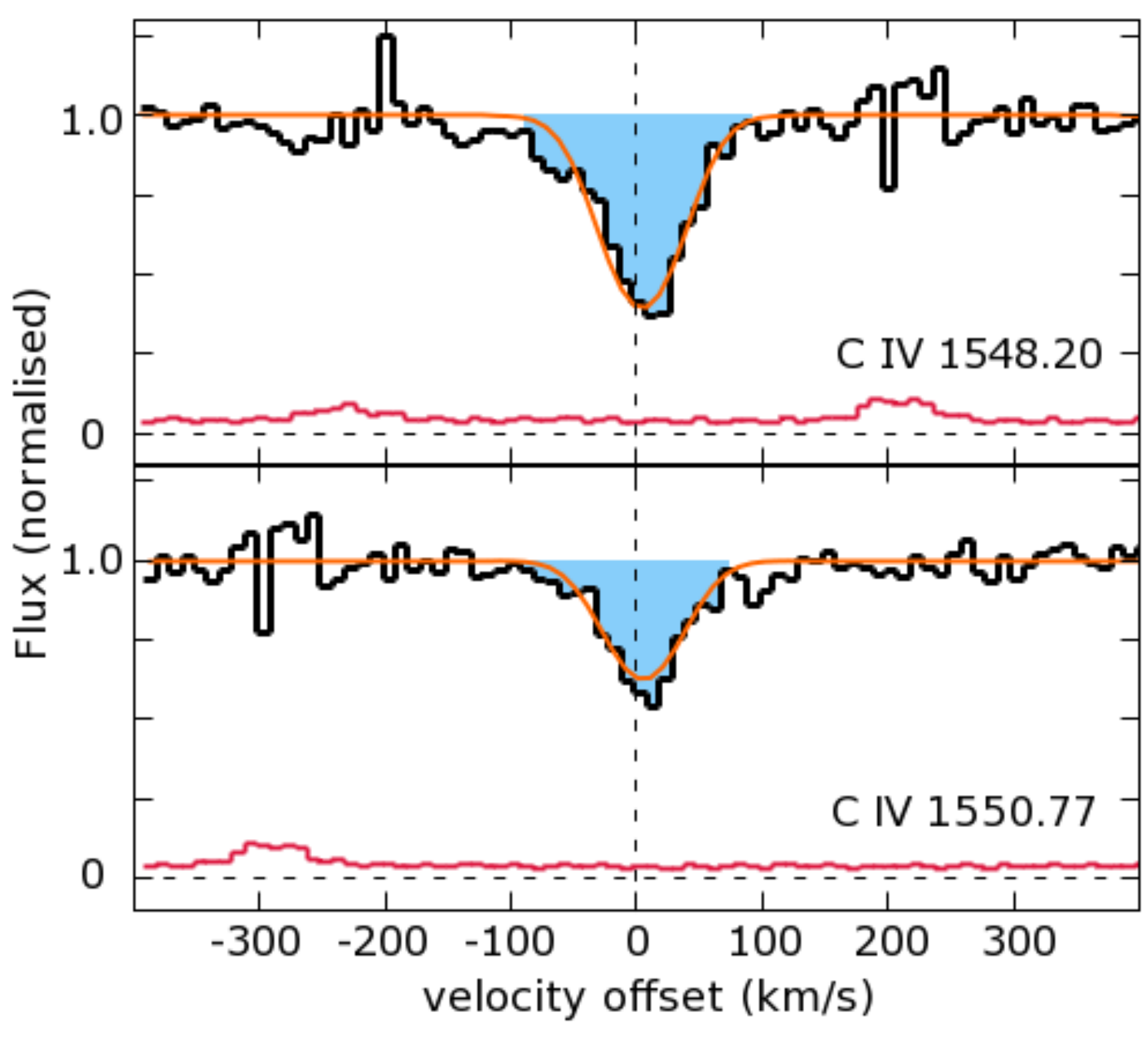}
\caption{Intervening system at $z=5.79539$.  Lines are as in Figure~\ref{fig:cfour_z6p51}.}
\end{figure}

\begin{figure}
\includegraphics[width=\columnwidth]{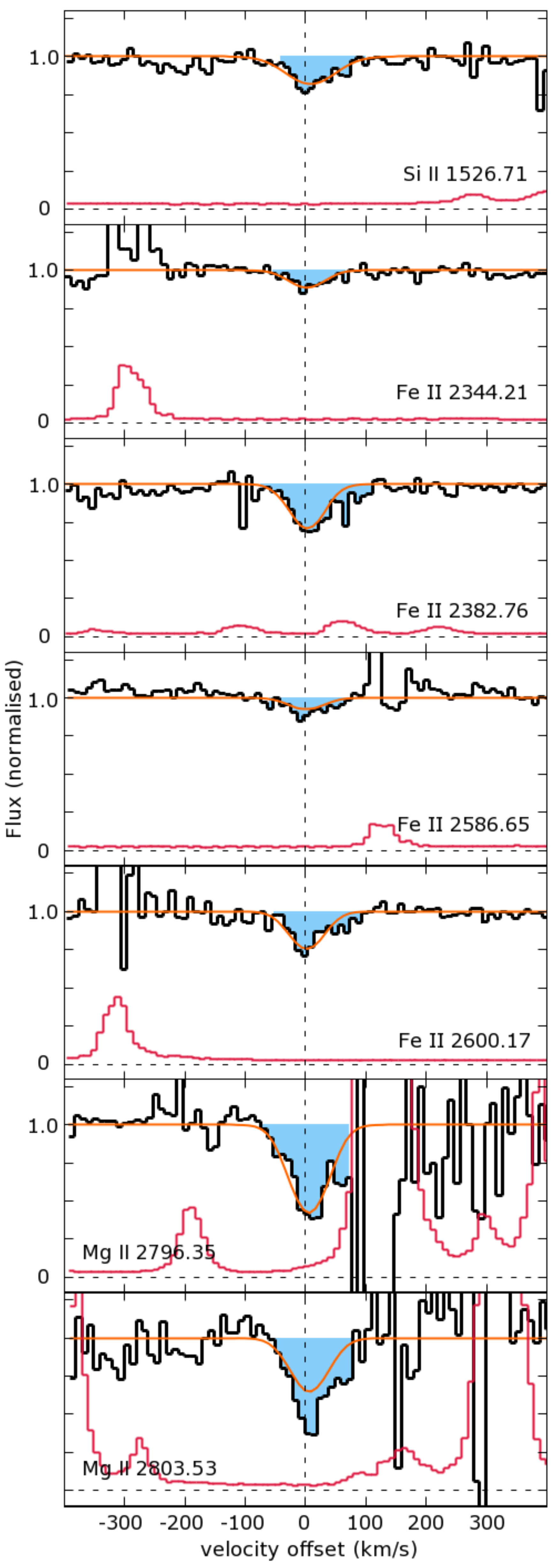}
\caption{Intervening system at $z=5.50793$.  Lines are as in Figure~\ref{fig:cfour_z6p51}.  Both transitions of \magtwo\ are strongly affected by skyline residuals at $\Delta v \gtrsim 70$ km s$^{-1}$. In this case, the column density of \magtwo is measured using the 2796.35 \AA\ transition alone.}
\end{figure}

\begin{figure}
\includegraphics[width=\columnwidth]{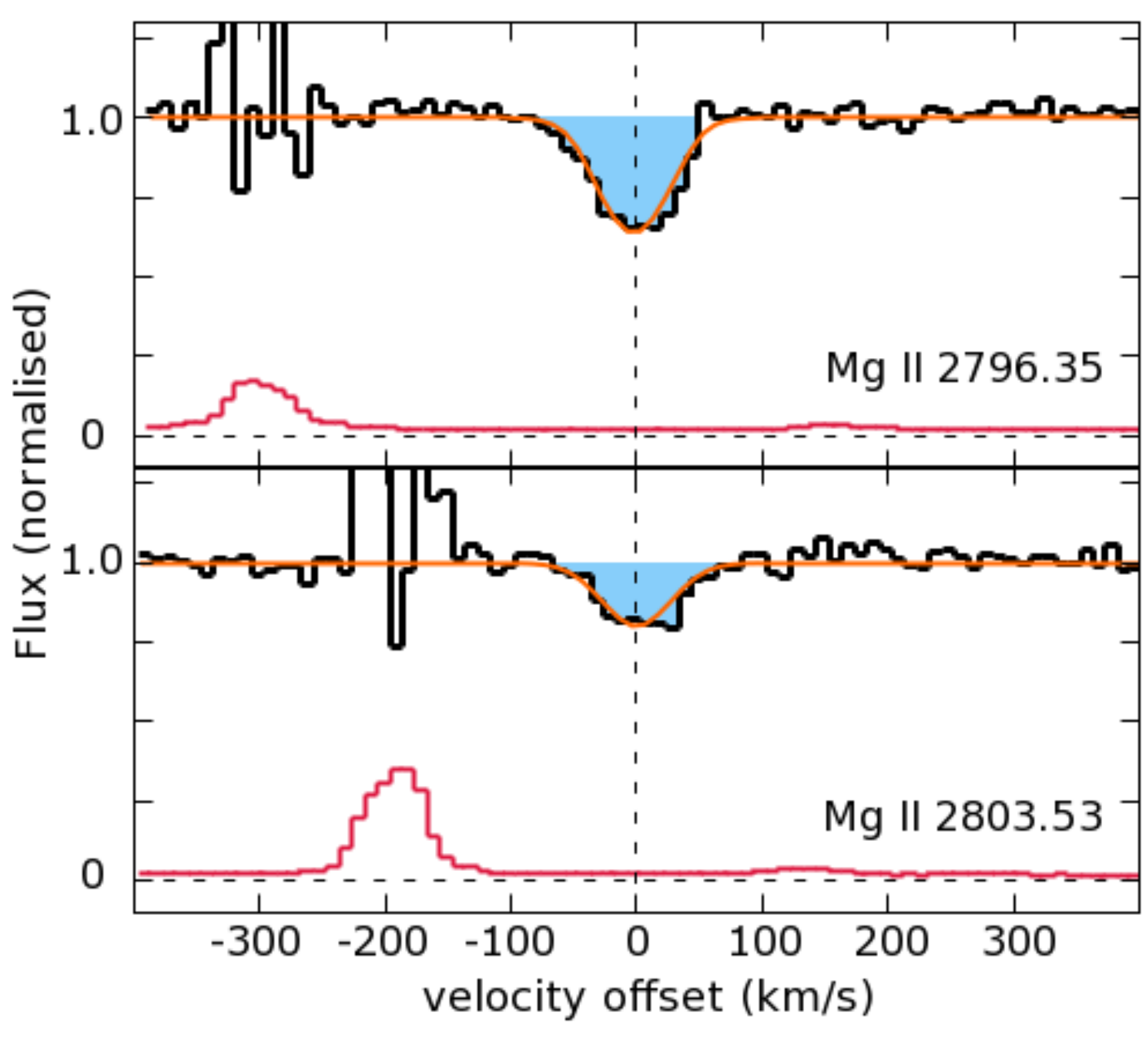}
\caption{Intervening system at $z=4.47260$.  Lines are as in Figure~\ref{fig:cfour_z6p51}.}
\end{figure}

\begin{figure}
\includegraphics[width=\columnwidth]{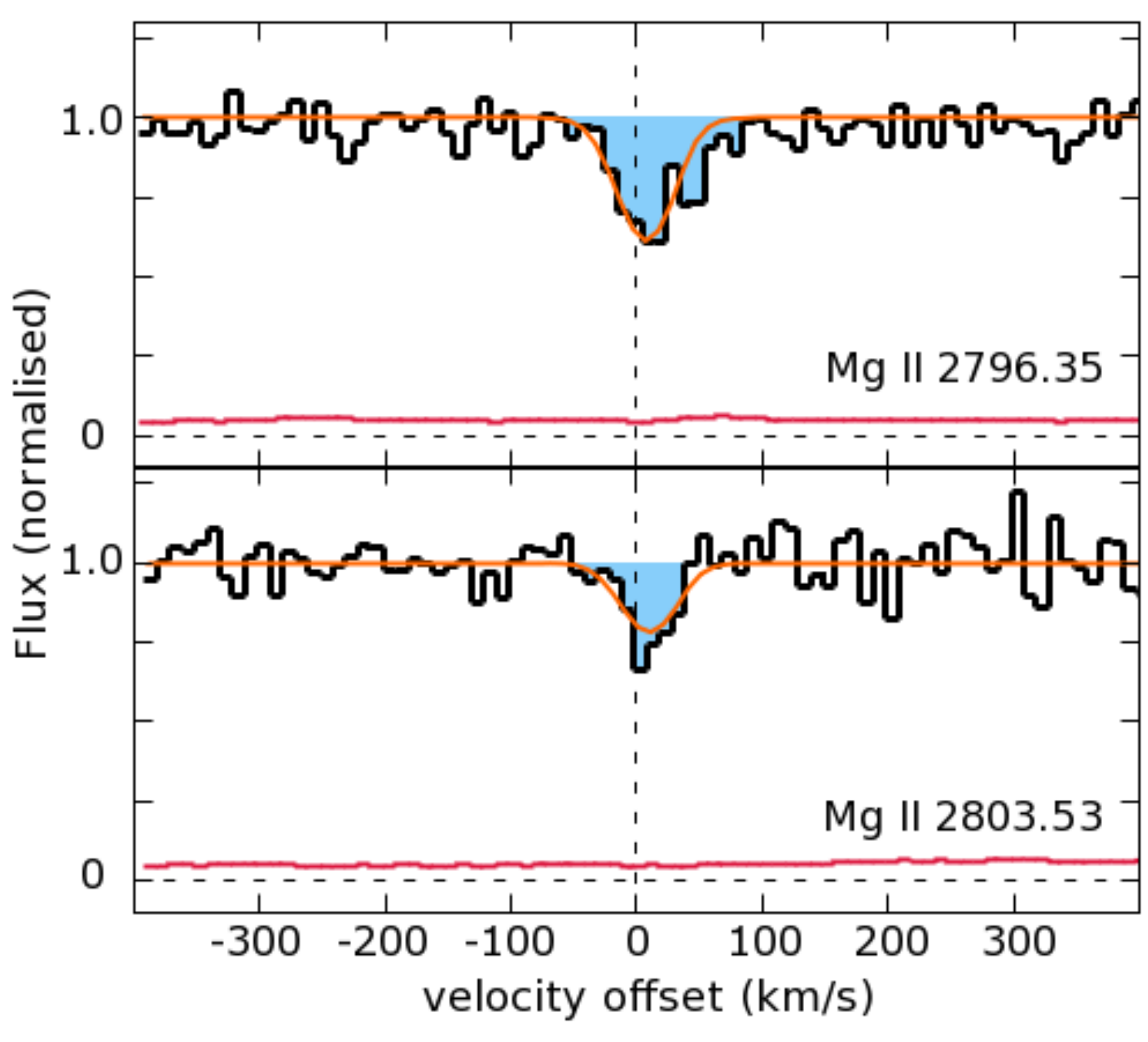}
\caption{Intervening system at $z=2.80961$.  Lines are as in Figure~\ref{fig:cfour_z6p51}.}
\end{figure}

\begin{figure}
\includegraphics[width=\columnwidth]{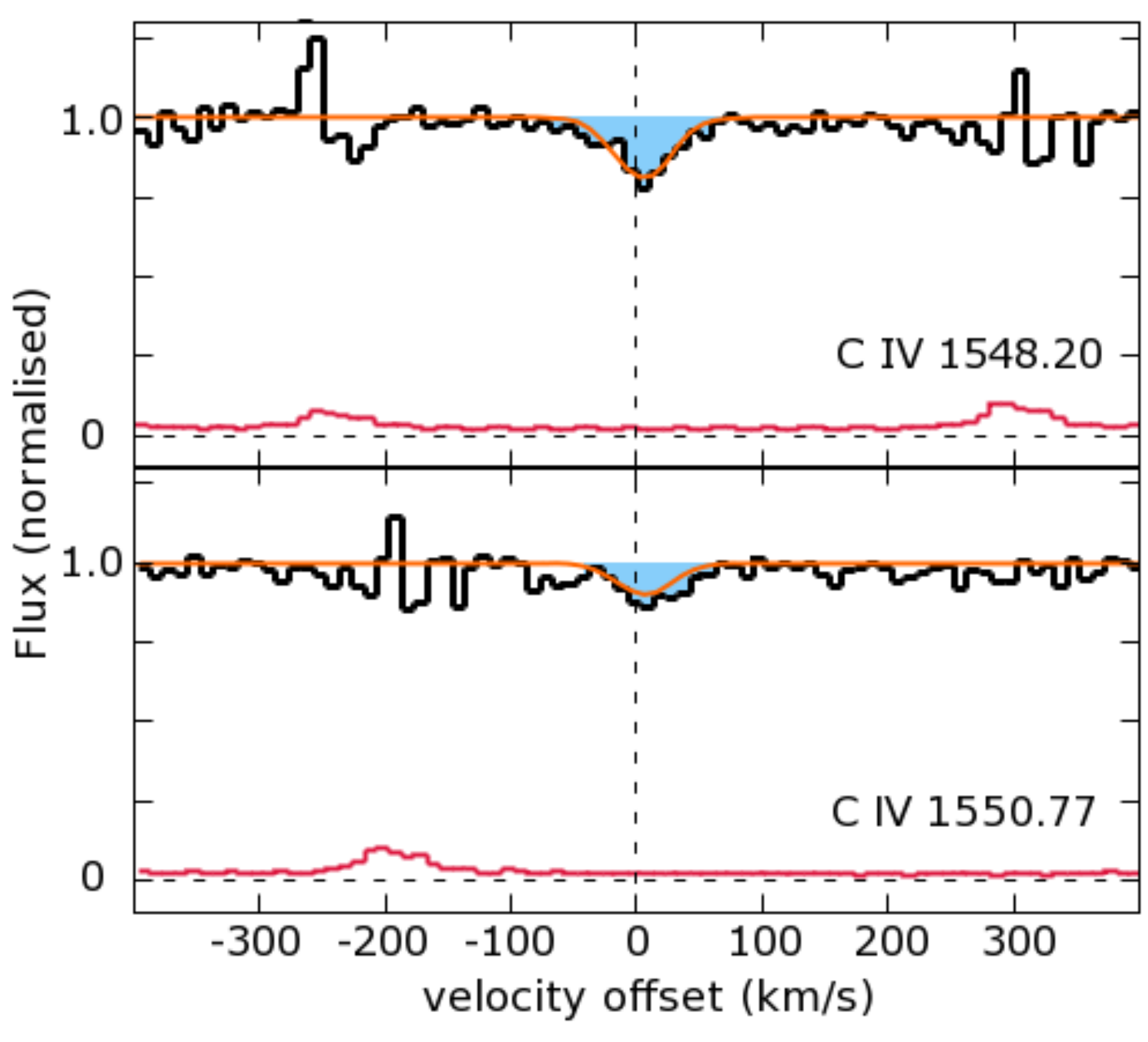}
\caption{Intervening or possibly associated system at $z=7.01652$.  Lines are as in Figure~\ref{fig:cfour_z6p51}.}
\label{fig:last}
\end{figure}

\end{document}